\documentclass[aps, prd, nofootinbib, twocolumn, preprintnumbers, showpacs, floatfix]{revtex4}
\usepackage{graphicx}
\usepackage{amsmath}
\usepackage{multirow}
\usepackage{bm}
\usepackage{color}
\def\fsl#1{\setbox0=\hbox{$#1$}                 
   \dimen0=\wd0                                 
   \setbox1=\hbox{/} \dimen1=\wd1               
   \ifdim\dimen0>\dimen1                        
      \rlap{\hbox to \dimen0{\hfil/\hfil}}      
      #1                                        
   \else                                        
      \rlap{\hbox to \dimen1{\hfil$#1$\hfil}}   
      /                                         
      \fi}                                      %

\newcommand{\VEV}[1]{\langle #1 \rangle}

\newcommand{\simg}{\hspace{0.3em}\raisebox{0.4ex}{$>$}\hspace{-0.75em}\raisebox{-.7ex}{$\sim$}\hspace{0.3em}}
\newcommand{\GB}{ \langle \alpha G_{\mu \nu}^2 \rangle}
\begin{document}
\title{Techni-dilaton at Conformal Edge \vspace{5mm}}
 \author{Michio Hashimoto}
 \affiliation{
  Maskawa Institute for Science and Culture, Kyoto Sangyo University, \\
  Motoyama, Kamigamo, Kita-Ku, Kyoto 603-8555, JAPAN.}
 \email{michioh@cc.kyoto-su.ac.jp}

 \author{Koichi Yamawaki}
  \affiliation{
  Kobayashi-Maskawa Institute for 
  the Origin of Particles and the Universe (KMI), \\
  Nagoya University, \\
  Nagoya 464-8602, JAPAN.}
 \email{yamawaki@kmi.nagoya-u.ac.jp}

\begin{abstract}
Techni-dilaton (TD)  was proposed long ago in the technicolor (TC) 
near criticality/conformality. To reveal the critical behavior of  TD,
we {\it explicitly} compute  the {\it nonperturbative} contributions 
to the scale anomaly $\langle \theta^\mu_\mu\rangle $ and 
to the techni-gluon condensate 
$\langle  \alpha G_{\mu\nu}^2 \rangle$, which are generated by
the dynamical mass $m$ of the techni-fermions. 
Our computation is  based on the (improved) 
ladder Schwinger-Dyson equation,  with the gauge coupling $\alpha$ 
replaced by the {\it  two-loop running coupling} $\alpha(\mu)$ having
the {\it Caswell-Banks-Zaks infrared fixed point} $\alpha_*$:
$\alpha(\mu) \simeq \alpha = \alpha_*$ for the infrared region 
$m < \mu < \Lambda_{\rm TC}$,  where $\Lambda_{\rm TC} $ is 
the intrinsic scale (analogue of  $\Lambda_{\rm QCD}$ of QCD)
relevant to the {\it perturbative} scale anomaly.
We find that
$-\langle \theta^\mu_\mu\rangle/m^4  \rightarrow {\rm const} \ne 0$ and
$\langle \alpha G_{\mu\nu}^2\rangle/m^4 \rightarrow 
 (\alpha/\alpha_{\rm cr}-1)^{-3/2} \rightarrow \infty$
in the criticality limit 
$m/\Lambda_{\rm TC} \sim 
 \exp (-\pi / (\alpha/\alpha_{\rm cr}-1)^{1/2})  \rightarrow 0$ 
($\alpha=\alpha_* \searrow \alpha_{\rm cr}$, or $N_f \nearrow N_f^{\rm cr}$) 
(``conformal edge''). 
Our result precisely reproduces the formal identity 
$\langle \theta^\mu_\mu\rangle =(\beta(\alpha) /4 \alpha^2)
 \langle  \alpha G_{\mu\nu}^2\rangle $, where 
$\beta(\alpha) = \Lambda_{\rm TC}\frac{\partial \alpha}{\partial \Lambda_{\rm TC}}
 = -(2\alpha_{\rm cr}/\pi) \cdot (\alpha/\alpha_{\rm cr}-1)^{3/2}$ is 
the {\it nonperturbative} beta function corresponding to 
the above essential singularity scaling of $m/\Lambda_{\rm TC}$. 
Accordingly,  the  PCDC (Partially Conserved Dilatation Current) implies
$(M_{\rm TD}/m)^2 (F_{\rm TD}/m)^2 = -4 \langle \theta_\mu^\mu\rangle/m^4 
 \rightarrow {\rm const} \ne 0$ at criticality limit,
where $M_{\rm TD}$ is the mass of TD and $F_{\rm TD}$ the decay constant of TD.
We thus conclude that at criticality limit 
the TD could become a ``{\it true (massless)  Nambu-Goldstone boson}'' 
$M_{\rm TD}/m\rightarrow 0$,  
only when $m/F_{\rm TD}\rightarrow 0$, namely  getting {\it decoupled}, 
as was the case of ``holographic techni-dilaton'' of Haba-Matsuzaki-Yamawaki.
The decoupled TD can be a candidate of dark matter.

\end{abstract}

\maketitle

\section{Introduction}

The conformal/scale-invariant (walking)  technicolor (TC) characterized 
by the large anomalous dimension $\gamma_m =1$ was first 
proposed~\cite{Yamawaki:1985zg,Bando:1986bg} as a solution to 
the problem of the Flavor-Changing Neutral Currents (FCNC) in TC,  
based  on the pioneering work by 
Maskawa and Nakajima~\cite{Maskawa:1974vs} who discovered 
{\it nonzero critical coupling}, $\alpha_{\rm cr} (\ne 0)$, 
for the  spontaneous chiral symmetry breaking (S$\chi$SB)  
to take place in  the ladder Schwinger-Dyson  (SD) equation 
with non-running (conformal) gauge coupling 
$\alpha(\mu) \equiv \alpha > \alpha_{\rm cr}$.~\footnote{
The solution of the FCNC problem  by the large anomalous dimension 
was first considered by B. Holdom~\cite{Holdom:1981rm}, 
based on a pure assumption  of  the existence of ultraviolet (UV) 
fixed point in TC without explicit dynamics and 
hence without definite prediction of the value of the anomalous dimension.
}
Subsequently, similar solution to FCNC within the same framework of 
the ladder SD equation was also considered without usage of 
the concept of the anomalous dimension~\cite{Akiba:1985rr}.
See for review \cite{Hill:2002ap}. 
 
Due to the (approximate) scale invariance, the theory also 
predicted~\cite{Yamawaki:1985zg,Bando:1986bg} a techni-dilaton (TD), 
a composite pseudo-Nambu-Goldstone (NG) boson for the spontaneous 
(and explicit) breaking of the scale symmetry of  the TC,  
as a techni-fermion and anti-techni-fermion bound state.  

Actually, the mass function of the fermion $\Sigma(Q)$ of 
the S$\chi$SB solution takes the asymptotic form~\cite{Maskawa:1974vs} 
\begin{eqnarray}
\Sigma (Q) \sim m^2/Q \quad (Q 
 \gg m) \,, 
\label{asymp}
\end{eqnarray}
which was interpreted as~\cite{Yamawaki:1985zg} 
\begin{equation}
\gamma_m=1,
\label{gammam}
\end{equation}
where the dynamical mass $m$  ($\Sigma(m)=m$) is given by
the form of essential singularity~\cite{Fukuda:1976zb, Miransky:1984ef}:
\begin{eqnarray}
 m &\sim& \Lambda\cdot
 \exp \left(-\frac{\pi}{\sqrt{\frac{\alpha}{\alpha_{\rm cr}} -1}}\right)
 \,,\quad
 \alpha_{\rm cr} = \frac{\pi}{3 C_F}\,,
\label{Miransky}
\end{eqnarray}
with $\Lambda$ being  the cutoff introduced to the SD equation  
and $C_F$ the quadratic Casimir of the fermion of 
the fundamental representation of the gauge group.  
 
\begin{figure}[t]
  \begin{center}
  \resizebox{0.4\textheight}{!}{\includegraphics{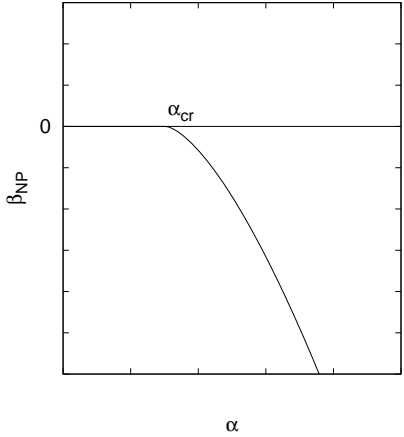}}
  \end{center}
  \caption{Schematic behavior of the nonperturbative $\beta(\alpha)$
   given in Eq.~(\ref{Miranskybeta}).
  \label{UVFP}}
\end{figure}

Eq.(\ref{Miransky}),  often called Miransky scaling,  
implies \cite{Miransky:1984ef} that the dynamical generation of 
$m$ by the nonperturbative dynamics  should lead to
the {\it nonperturbative} running of the coupling  
$\alpha=\alpha(\Lambda/m) $ ($\rightarrow \alpha_{\rm cr}$ as 
$\Lambda/m \rightarrow \infty$) even when it is nonrunning (conformal)  
in the perturbative sense: 
\begin{equation}
\beta_{_{\rm NP}} (\alpha) =
\Lambda \frac{\partial \alpha}{\partial\Lambda} = 
-\frac{2\alpha_{\rm cr}}{\pi} 
 \left(\frac{\alpha}{\alpha_{\rm cr}}-1\right )^{\frac{3}{2}} \,,
\label{Miranskybeta}
\end{equation}
with $\alpha_{\rm cr}$ being interpreted as the UV fixed point. 
See Fig.~\ref{UVFP}~\cite{Yamawaki:1985zg}. 
Actually, the mass scale of $m$ has {\it never been created from nothing} 
but transferred from the ``hidden scale'' $\Lambda$ whose effect
persists even when it is removed by taking the limit 
$\Lambda\rightarrow \infty$ while tuning $\alpha \rightarrow \alpha_{\rm cr}$.

It was then argued~\cite{Bando:1986bg} that 
the dynamically generated mass $m$ is 
a renormalization-group (RG) -- invariant quantity,
$\frac{d m}{d \mu} =0$,  
and  is regarded as generated by the {\it dimensional transmutation}: 
\begin{equation}
  m = \mu \cdot \exp 
  \left(  - \int^{\alpha(\mu)} \frac{d \alpha}{\beta_{_{\rm NP}}(\alpha) }\right),
\label{dimtrans}
\end{equation} 
due to (nonperturbative) running of $\alpha(\mu)$,  
such that 
$\mu \frac{\partial \alpha(\mu)}{\partial \mu}=\beta_{_{\rm NP}}(\alpha(\mu))$, 
which reflects the (nonperturbative) scale anomaly  
\begin{equation}
  \langle \partial^\mu D_\mu\rangle = 
  \langle\theta^\mu_\mu \rangle = 
  \frac{\beta_{_{\rm NP}}(\alpha)}{4 \alpha^2} \GB\, ,  
\label{anomaly}
\end{equation} 
with $\beta_{_{\rm NP}}(\alpha)$ given in Eq.(\ref{Miranskybeta}),  
where $\GB$ is the {\it nonperturbative} contribution to 
the techni-gluon condensate  due to the mass generation of $m$. 
Note that the non-perturbative beta function (\ref{Miranskybeta})  
has a {\it multiple zero}, i.e., 
$\beta \sim - (\alpha/\alpha_{\rm cr}-1)^\delta$ with $\delta > 1$, 
which is crucial for Eq. (\ref{dimtrans}) 
with Eq.~(\ref{Miranskybeta}) to reproduce the essential singularity scaling
Eq.(\ref{Miransky}) for $\mu=\Lambda$.  

Initially  it was assumed~\cite{Yamawaki:1985zg,Bando:1986bg} that 
\begin{equation}
 \GB = {\cal O} (m^4) \, ,
\label{false}
\end{equation}
so that
\begin{equation}
 \langle \theta_\mu^\mu\rangle =
 \beta_{_{\rm NP}} (\alpha) \cdot  {\cal O} (m^4) \,
\label{vacenergy0}
\end{equation}
namely $\VEV{\theta_\mu^\mu}/m^4 \sim \beta_{_{\rm  NP}}(\alpha) \rightarrow 0$  
in the criticality limit $\alpha\rightarrow \alpha_{\rm cr}$. From the scale 
anomaly through Partially Conserved Dilatation Currents (PCDC) ,  
the TD mass ($M_{\rm TD}$) and its decay constant ($F_{\rm TD}$)
are related as  
\begin{equation}
  M_{\rm TD}^2 F_{\rm TD}^2 =
  -4\langle \theta^\mu_\mu \rangle =
  - \frac{\beta_{_{\rm NP}}(\alpha)}{\alpha^2} \GB\, ,
\label{PCDC}
\end{equation}
which would imply 
$(M_{\rm TD}/m)^2 (F_{\rm TD}/m)^2 \sim - \beta_{_{\rm NP}}(\alpha)\rightarrow 0$
for the criticality $\alpha \rightarrow \alpha_{\rm cr}$.
It was then argued~\cite{Bando:1986bg} that  $M_{\rm TD}/m$ could be 
arbitrarily small by tuning $\alpha$ as  
$\beta_{_{\rm NP}}(\alpha) \rightarrow 0$ ($\alpha \rightarrow \alpha_{\rm cr}$), 
namely,  TD  could become a true Nambu-Goldstone (NG) boson, 
$M_{\rm TD}/m\rightarrow 0$,  in the criticality limit 
$\alpha \rightarrow \alpha_{\rm cr}$ ($m/\Lambda \rightarrow 0$).
 
Actually, Eq. (\ref{vacenergy0}) (and hence Eq.(\ref{false})) 
turned out to be false at least in the ladder approximation: 
In the criticality limit $\alpha \rightarrow \alpha_{\rm cr}$ 
($m/\Lambda \rightarrow 0$)  the straightforward  
ladder calculation~\cite{Miransky:1989qc} 
of  $\langle \theta^\mu_\mu\rangle= 4 \langle \theta^0_0\rangle$,  
with the vacuum energy 
$\langle \theta^0_0\rangle$ evaluated through the 
Cornwall--Jackiw--Tomboulis (CJT) effective potential,  yields
\begin{equation}
 \langle \theta^\mu_\mu\rangle  =-4 \frac{N_f N_{\rm TC}}{\pi^4} m^4\,,   
 \label{vacuumenergy}
\end{equation}
in obvious contradiction with Eq.(\ref{vacenergy0}).  Accordingly, we have 
\begin{eqnarray}
  \frac{\GB}{m^4} = \frac{4\alpha^2 }{\beta_{_{\rm NP}}(\alpha)}
  \cdot  \frac{\langle \theta^\mu_\mu\rangle}{m^4} 
  \sim -\frac{1}{\beta_{_{\rm NP}}(\alpha)} \rightarrow \infty\,,
  \label{divGB}
\end{eqnarray}
in contrast to Eq.(\ref{false}).  Eq.(\ref{PCDC}) now reads
\begin{equation}
\left( \frac{M_{\rm TD}}{m} \right)^2 \left(\frac{F_{\rm TD}}{m}\right)^2 =
  -\frac{4\langle \theta^\mu_\mu \rangle}{m^4} \rightarrow
   {\rm const} \ne 0\,,
\label{PCDC2}
\end{equation}
which implies that there is no massless TD in the criticality limit, 
$M_{\rm TD}/m  \rightarrow {\rm const}\ne 0$,  as far as 
$m/F_{\rm TD} \rightarrow {\rm const} \ne 0$. ~\footnote{
There were several other arguments against the massless dilaton in 
the criticality limit of the (nearly) conformal/scale-invariant 
gauge theories in ladder-type approximation~\cite{Bardeen:1985sm,Holdom:1986ub,
Kondo:1988qd,Nonoyama:1989dq}.
}

Recently a possibility was suggested~\cite{Yamawaki:2009vb} that 
the TD is relatively light compared with other techni-hadrons, 
though not extremely light:
The TD mass may be evaluated at $\alpha=\alpha_{\rm cr}$ in the limit 
$m/\Lambda \rightarrow 0$ as 
\begin{equation}
M_{\rm TD} \simeq \sqrt{2} \, m\,,
\label{gaugedNJLTD}
\end{equation}
through the old calculation~\cite{Shuto:1989te}  of a scalar bound state 
in the gauged NJL model which well describes 
the conformal/scale-invariant gauge dynamics  at criticality 
$\alpha \rightarrow  \alpha_{\rm cr}$ where the anomalous dimension 
$\gamma_m =1$ makes the induced four-fermion operator marginal
with physical dimension $d=2(3-\gamma_m)=4$~\cite{Bardeen:1985sm}.  
Eq.(\ref{gaugedNJLTD}) is consistent with  the ladder calculation  
Eq.(\ref{vacuumenergy}) (and hence Eq.(\ref{PCDC2})).

Furthermore, in the modern 
version~\cite{Lane:1991qh,Appelquist:1996dq, Miransky:1996pd}  
of  conformal/scale-invariant TC based on 
the Caswell-Banks-Zaks infrared fixed point (CBZ-IRFP), 
$\alpha_* =\alpha_*(N_f,N_{\rm TC})$~\cite{Caswell:1974gg}, of  
the two-loop beta function, 
the coupling is almost nonrunning $\alpha(\mu) \simeq \alpha=\alpha_*$ 
over the wide infrared (IR) region $\mu<\Lambda_{\rm TC}$ below 
the intrinsic scale $\Lambda_{\rm TC}$ 
which is an analogue of $\Lambda_{\rm QCD}$ of QCD (see Eq.(\ref{LambdaTC})).
Based on the SD equation in the (improved) ladder approximation, 
with the nonrunning coupling $\alpha$ in the ladder expression 
simply replaced by the two-loop running coupling $\alpha(\mu)$,
we have approximately the same result as 
Eq.(\ref{asymp})--Eq.(\ref{Miransky}) with the cutoff $\Lambda$ 
replaced by $\Lambda_{\rm TC}$ (to be typically identified with 
the Extended TC scale $\Lambda_{\rm ETC}$), 
and the nonperturbative beta function Eq.(\ref{Miranskybeta})
as well as $\gamma_m \simeq 1$ near criticality.
 
In this case  such a relatively light TD was also 
suggested~\cite{Yamawaki:2007zz,Yamawaki:2009vb}  from 
the result of  the straightforward calculation~\cite{Harada:2003dc} 
of scalar bound state mass,
\begin{eqnarray}
 M_{\rm TD}  &\simeq&  1.5\, m \,(\simeq 4 F_\pi)  \simeq \sqrt{2} \,m 
 \label{SDBS} \\
\nonumber
 &<& \, M_\rho, M_{a_1} \simeq 4.2 \, m\, ,
\end{eqnarray}
near the criticality $\alpha_*\simeq \alpha_{\rm cr} \;(N_f \simeq N_f^{\rm cr})$,
where $N_f^{\rm cr}=N_f^{\rm cr}(N_{\rm TC})$ is determined by 
$\alpha_*=\alpha_{\rm cr}$~\cite{Appelquist:1996dq}
and $M_\rho, M_{a_1} $ are mass of techni-$\rho$ and techni-$a_1$ mesons. 
The calculations are based on the SD equation and the homogeneous 
Bethe-Salpeter (BS) equation in the  improved ladder approximation.
~\footnote{
These results are compared with those in QCD: ~\cite{Kurachi:2006ej}
$(M_\rho/F_\pi)/(M_\rho/F_\pi)_{\rm QCD}\simeq 1.3$, 
$(M_{a_1}/F_\pi)/(M_{a_1}/F_\pi)_{\rm QCD} \simeq 0.86$, while
$(M_{\rm TD}/F_\pi)/(M_{\rm scalar}/F_\pi)_{\rm QCD}\simeq 0.38$.
}
Although the result Eq.(\ref{SDBS}) is evaluated not in 
the critical limit $\alpha=\alpha_* \rightarrow \alpha_{\rm cr}$ 
but slightly away from it,  the result seems to indicate
$M_{\rm TD}/m \rightarrow {\rm const}\ne 0$ in the criticality limit 
$\alpha=\alpha_* \rightarrow \alpha_{\rm cr}$,  
which is consistent with the ladder calculation Eq.(\ref{PCDC2}).
Numerically, Eq.(\ref{SDBS}) suggests~\cite{Yamawaki:2007zz,Yamawaki:2009vb} 
\begin{equation}
M_{\rm TD} \sim 500 \, {\rm GeV}
\label{TD1}
\end{equation}
in the typical one-family TC model near criticality 
(with  $N_f \simeq 4N_{\rm TC} =$ 8--12).

More recently, TD mass was estimated by Haba, 
Matsuzaki and Yamawaki~\cite{Haba:2010iv}
in the hard-wall type (bottom-up) holographic approach including 
the techni-gluon condensate $\GB$.
It was found that 
for fixed value of  $S\,$ and $\gamma_m$, \footnote{
Note that 
in the holography the $S$ parameter~\cite{PeskinTakeuchi} and 
anomalous dimension $\gamma_m$ 
are not  calculable parameters but 
arbitrary adjustable parameters~\cite{Haba:2008nz}. 
} 
the TD mass is a monotonically decreasing function 
\begin{equation}
\frac{M_{\rm TD}}{m}  \rightarrow 0\quad (\Gamma \rightarrow \infty)\,, 
\label{NGlimit}
\end{equation}
of  the techni-gluon condensate  $\Gamma$ 
(normalized by the corresponding quantity of QCD), 
\begin{equation}
\label{Gamma}
\Gamma 
\equiv \left(
\frac{\GB/F_\pi^4}
{\left(\GB/F_\pi^4\right)_{\rm QCD}} 
\right)^{1/4} \,,  
\end{equation}
 where $F_\pi$ is the decay constant of the (techni-) pion of order 
${\cal O} (m)$.
It was argued that the limit $\Gamma \rightarrow \infty$ is realized at the criticality 
$\beta_{_{\rm NP}}(\alpha) \rightarrow 0$ 
($\alpha=\alpha_* \rightarrow \alpha_{\rm cr}$), 
as is seen from Eq. (\ref{divGB}),   when the value of $\GB$ is evaluated  through Eq.(\ref{anomaly})
by assuming the ladder result  Eq.(\ref{vacuumenergy}) and 
the nonperturbative beta function Eq.(\ref{Miranskybeta}). 
This  would imply  the existence of {\it true (massless) NG boson} 
at criticality in contrast to Eq.(\ref{gaugedNJLTD}) and (possibly) 
Eq.(\ref{SDBS}).  However, from the ladder result  Eq. (\ref{PCDC2}), 
Eq. (\ref{NGlimit}) implies that
\begin{equation}
\frac{m}{F_{\rm TD}} \sim \frac{M_{\rm TD}}{m} \rightarrow 0
\label{decoupledTD}
\end{equation}
in that limit, namely the holographic TD becomes a {\it decoupled TD} 
whose all couplings are characterized by the power of 
$(p/F_{\rm TD})$ with the typical momentum $p (\sim m)$.
This is a new feature of the holographic TD.
 
The actual phenomenologically interesting situation of  
TC model building  is slightly away from the criticality,  
$m/\Lambda=m/\Lambda_{\rm ETC}\simeq 10^{-3}-10^{-4} \ne 0$, \footnote{
In the actual situation of TC, $m$ is the weak scale 
$m \sim {\rm TeV}$ and  $\Lambda$  is identified with 
the typical scale $\Lambda_{\rm ETC}$ of the dynamics 
(like the extended TC (ETC)) transmitting the techni-fermion mass $m$ 
to that of the quark/lepton, i.e., 
$\Lambda=\Lambda_{\rm ETC}\sim 10^{3} {\rm TeV}$. 
Thus $m/\Lambda \sim 10^{-3}-10^{-4}$ which corresponds to  
$\beta(\alpha) \simg 10^{-2}$ 
(for $N_{\rm TC} = 2-3$)  from Eq.(\ref{Miransky}) and Eq.(\ref{Miranskybeta}).
}
in which case we have $\Gamma \sim 7$. 
This implies mass of holographic TD for
typical conformal/scale-invariant TC model with $N_f \simeq 4 N_{\rm TC}$ 
as:~\cite{Haba:2010iv}
\begin{eqnarray}
M_{\rm TD} &\simeq& 600 \,\, {\rm GeV}\, ,\\
\nonumber
&<& M_\rho, M_{a_1} \simeq  3. 8 \, \, {\rm TeV}
\label{TD2}
\end{eqnarray}
for the value of $S =0.1$ and $\gamma_m=1$, in  rough agreement 
with  Eq. (\ref{TD1}).~\footnote{
The value $M_\rho, M_{a_1}$ is essentially determined by the value of 
$S$: Lower $S$ value corresponds to higher $M_\rho,M_{a_1}$.
The calculated $S$ value in  Ref.~\cite{Harada:2005ru} in  
the same setting as Ref.~\cite{Harada:2003dc}  is higher than $S=0.1$,  
which corresponds to the value of $M_\rho, M_{a_1}$ in the holography close to
that of  Eq.(\ref{SDBS}).
}
 
Most recently, on the other hand, Appelquist and 
Bai~\cite{Appelquist:2010gy} argued,
based on the improved ladder SD equation with the two-loop running coupling, 
that there does exist a (non-decoupled) massless TD, 
$M_{\rm TD}/m \rightarrow 0$,  
in the conformal/scale-invariant TC in the  criticality limit 
$\beta(\alpha) \rightarrow 0$
as $\alpha=\alpha_*\rightarrow  \alpha_{\rm cr}\,
(N_f \rightarrow  N_f^{\rm cr})$,  based on essentially 
the same assumption as in  Ref.~\cite{Yamawaki:1985zg,Bando:1986bg},  
namely Eq.(\ref{false}), which is in disagreement with the ladder
calculation, Eq.(\ref{vacuumenergy}),  as noted before. (See also Ref.~\cite{Dietrich:2005jn}.)
Note that although 
the beta function in Ref.~\cite{Appelquist:2010gy} is 
somewhat different from that in Eq. (\ref{Miranskybeta}) used in 
Ref.~\cite{Bando:1986bg},  
they both vanish at the criticality 
$\alpha_* \rightarrow \alpha_{\rm cr}$ ($N_f\rightarrow N_f^{\rm cr}$).

In view of these subtleties in the literature on the critical behavior 
of the TD near the conformal edge associated with the CBZ-IRFP,
it is very  important to settle the  critical behavior of $\GB$ and 
$\langle \theta_\mu^\mu\rangle$
in the calculation within the same framework as that relevant to
the above controversy, namely, literally 
{\it incorporating  the perturbative two-loop running effects} 
as well the nonperturbative effects which produce the dynamical mass $m$. 

In this paper we shall explicitly calculate the nonperturbative 
contributions to the techni-gluon condensate $\GB$ and 
to the scale anomaly 
$\langle \theta_\mu^\mu\rangle=4 \langle \theta_0^0\rangle$ 
arising from the fermion mass generation 
in the TC near conformality/criticality (conformal edge), 
$\alpha_*\rightarrow  \alpha_{\rm cr} \, (N_f \rightarrow N_f^{\rm cr})$,
based on  the ``improved ladder SD equation''~\cite{Miransky:1983vj}.
Although the improved ladder approximation with the two-loop running 
coupling as well as the ladder approximation with nonrunning coupling 
is not a systematic approximation and hence not very reliable, 
all the above controversy about the techni-dilaton 
in the literature has been confined to this approximation. 
So our aim of this paper is to resolve the confusion within 
this approximation. 
We first study analytically the solution of  
the improved ladder SD equation, with the two-loop running coupling 
being approximated by a simplified ansatz 
(solution to the ``parabolic'' beta function
$\beta (\alpha) = - b_0 \alpha (\alpha_* - \alpha)$), 
\begin{equation}
  \alpha (\mu^2) =
  \frac{\alpha_*}
       {1+e^{-1}\left(\frac{\mu^2}{\Lambda_{\rm TC}^2}\right)^{b_0 \alpha_*}},
\end{equation}
which agrees with the exact two-loop running coupling written 
in terms of the Lambert's $W$ function in the IR region 
$\mu <\Lambda_{\rm TC}$ responsible for the dynamical mass 
generation (see text).
The result will be checked by the numerical solution 
based on the exact two-loop running coupling. 

We then calculate the techni-gluon condensate near the conformal edge
and show explicitly it behaves as
$\VEV{G_{\mu\nu}^2}/m^4 \sim (\alpha/\alpha_{\rm cr}-1)^{-3/2} \to \infty$
($\alpha \to \alpha_{\rm cr}$),  which is a direct evidence against  
the assumption of Ref.~\cite{Bando:1986bg} 
(in the ladder SD equation with nonrunning coupling) and 
also that of Ref.~\cite{Appelquist:2010gy}
(in the improved ladder SD equation with the two-loop running coupling). 
Our result directly confirms the estimate of the techni-gluon condensate 
in Ref.~\cite{Haba:2010iv} which indicates divergence of 
the techni-gluon condensate $\Gamma \rightarrow \infty$ at criticality.

We also find that the numerical calculation of 
the vacuum energy with the two-loop running coupling agrees with 
the analytical solution (\ref{vacuumenergy}) with the fixed coupling,
$\VEV{\theta_\mu^\mu} \sim -m^4$,  again in contrast to the assumption in 
Ref.~\cite{Bando:1986bg} and Ref.~\cite{Appelquist:2010gy}.

On the other hand, the scale anomaly satisfies the formal relation,
$\VEV{\theta_\mu^\mu}=\beta(\alpha)/(4\alpha)\cdot \VEV{G_{\mu\nu}^2}$. Hence
our results imply  the beta function near criticality:
\begin{equation}
\beta(\alpha)= \frac{4\alpha \VEV{\theta_\mu^\mu}}{\VEV{G_{\mu\nu}^2}}
 \sim - \left(\frac{\alpha}{\alpha_{\rm cr}}-1\right)^{\frac{3}{2}}\, .
\end{equation} 
The result also confirms the assumption made in Ref.~\cite{Haba:2010iv} 
where the nonperturbative beta function, Eq.~(\ref{Miranskybeta}), 
as well as the ladder result of the vacuum energy was used 
for the nonperturbative conformal anomaly to estimate 
the techni-gluon condensate.

We thus conclude that the nonperturbative beta function arising from 
the nonperturbative effects of the dynamical mass generation 
in the IR region $(\mu < \Lambda_{\rm TC})$ is 
essentially like Eq.~(\ref{Miranskybeta}),   Fig.\ref{UVFP},  
even in the case 
of the two-loop running gauge coupling set in the SD equation. 
It should be considerably changed from the perturbative expression
near criticality.
In Fig.~\ref{fig-beta-conj},
we schematically depict the conjectured behavior of 
the beta function including both of the perturbative and 
nonperturbative region.

Our two-loop results combined with 
the PCDC relation, 
$F_{\rm TD}^2 M_{\rm TD}^2 = - 4 \VEV{\theta_\mu^\mu} \sim m^4$,
suggest $F_{\rm TD}^2/m^2 \cdot M_{\rm TD}^2/m^2 \to $ finite
at the critical point, which is the same as 
the nonrunning case (\ref{PCDC2});
There is no theoretically controllable suppression factor 
for $M_{\rm TD}/m \to 0$, as far as $F_{\rm TD}/m$ is finite.
This contradicts the assumptions in Ref.~\cite{Bando:1986bg}
and Ref.~\cite{Appelquist:2010gy}. 
However, our results cannot exclude the possibility 
(the ``decoupled TD'', Eqs.(\ref{NGlimit}) and (\ref{decoupledTD})) that 
there might exist a very light TD, $M_{\rm TD} \sim 0$,
if $F_{\rm TD}/m$ is quite large,  
as could be the case in the  limit of the holographic TD~\cite{Haba:2010iv}.
This decoupled TD may be dark matter.

This paper is organized as follows:
In Sec.~II, we describe the behavior of the beta function
in the two-loop approximation.
We also introduce the parabolic approximation
in order to solve analytically the improved ladder SD equation.
In Sec.III, we study the analytical solution of the SD equation
in the parabolic approximation and also analyze 
the numerical solution with the two-loop exact gauge coupling.
We show that the approximation works well.
Then we calculate the techni-gluon condensate and the vacuum energy.
Sec.~IV is devoted for summary and discussions.

\section{Two-loop $\beta$ function and parabolic approximation}

In this section, we study the running effect of the gauge coupling 
constant in the two-loop approximation.
It is well-known that there appears the CBZ-IRFP~\cite{Caswell:1974gg},
when the number of (techni-)fermions is in a certain range, 
as we will show later.
If the value of the CBZ-IRFP $\alpha_*$ slightly exceeds 
the critical coupling $\alpha_{\rm cr}$ for the S$\chi$SB,  
we can apply such gauge theories to the TC with near conformality 
with anomalous dimension 
$\gamma_m \simeq 1$~\cite{Lane:1991qh,Appelquist:1996dq, Miransky:1996pd}.
We here employ the approach of 
the (improved) ladder SD equation~\cite{Miransky:1983vj}, 
with the nonrunning coupling in the ladder SD equation simply replaced by
the running one,  this time the two-loop running coupling.
Although the numerical analysis of the (improved) ladder SD equation is
rather straightforward, it is not so easy to extract numerically 
the critical behavior of the solution.
We thus approximate the two-loop $\beta$ function
into a parabolic one and will solve {\it analytically} 
the ladder SD equation.
In the next section, 
we will demonstrate that this approximation works very well

Let us study the two-loop renormalization group equation (RGE) 
for the gauge coupling constant $\alpha$~\cite{pdg}:
\begin{equation}
  \mu \frac{\partial }{\partial \mu}\alpha = \beta (\alpha)=
  -b_0 \alpha^2 -b_1 \alpha^3, 
  \label{banks-zaks}
\end{equation}
with
\begin{equation}
  b_0 = \frac{1}{6\pi}(11C_A - 4 N_f T_R), 
\end{equation}
and
\begin{equation}
    b_1 = \frac{1}{12\pi^2}\bigg[\,17C_A^2 - 2 N_f T_R(5C_A+3C_F)\,\bigg], 
\end{equation}
where $N_f$ represents the number of flavor and
the group theoretical factors are 
\begin{equation}
  C_A=N_{\rm TC}, \quad T_R = \frac{1}{2}, \quad 
  C_F = \frac{N_{\rm TC}^2-1}{2N_{\rm TC}},
\end{equation}
for $SU(N_{\rm  TC})$ gauge theories.

\begin{figure}[t]
  \begin{center}
  \resizebox{0.4\textheight}{!}{\includegraphics{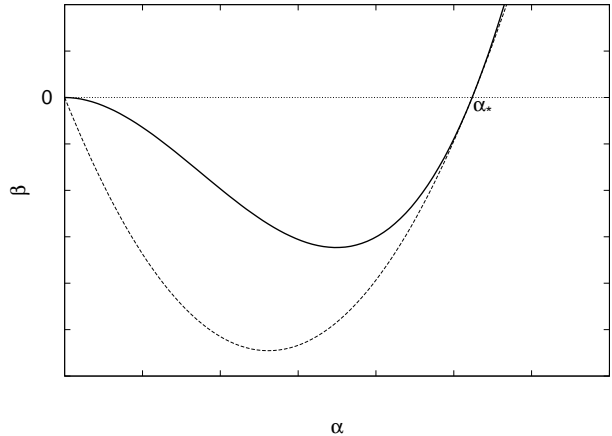}}
  \end{center}
  \caption{Behavior of $\beta(\alpha)$ in perturbation.
    The bold solid and dashed curves correspond to
    the two-loop $\beta$ function (\ref{banks-zaks}) and
    the parabolic one (\ref{app-parabolic}), respectively.
  \label{fig-beta}}
\end{figure}

When $b_0 > 0$ and $b_1 < 0$, i.e.,
\begin{equation}
  \frac{34 N_{\rm TC}^3}{13 N_{\rm TC}^2 -3} < N_f < \frac{11}{2}N_{\rm TC},
\end{equation}
the CBZ-IRFP $\alpha_*$ emerges,
\begin{equation}
  \alpha_* = \frac{b_0}{-b_1} \, .
\end{equation}
The analytic form of $\alpha(\mu^2)$ is 
also known~\cite{Gardi:1998rf}:
\begin{equation}
  \alpha (\mu^2) = \frac{\alpha_*}{1+W(z(\mu^2))},
  \label{alp-lambert}
\end{equation}
where $W$ denotes the Lambert function~\cite{Corless:1996zz}, 
which is the inverse of $x e^x$,
and $z$ is defined by
\begin{equation}
  z(\mu^2) \equiv
  \frac{1}{e}\left(\frac{\mu^2}{\Lambda_{\rm TC}^2}\right)^{b_0 \alpha_*},
  \label{z-lam0}
\end{equation}
where the  intrinsic scale $\Lambda_{\rm TC}$ analogous to 
$\Lambda_{\rm QCD}$ invariant under RGE  is given by~\cite{Appelquist:1996dq}:
\begin{eqnarray}
&& \hspace*{-4mm} 
   \Lambda_{\rm TC} = 
   \mu \cdot\exp \left(\int^{\alpha(\mu)} \frac{d\alpha}{\beta(\alpha)}\right) 
   \nonumber \\
&& = \mu \cdot\exp\Bigg[\,
  -\frac{1}{b_0 \alpha(\mu^2)}
-\frac{1}{b_0 \alpha_*}\ln \left(
   \frac{\alpha_* 
   -\alpha(\mu^2)}{\alpha(\mu^2)}\right)
\,\Bigg] \, ,    \nonumber \\ 
 \label{LambdaTC}
\end{eqnarray}
with the first term in $[\,...\,]$ being 
the usual one-loop contribution and the second the two-loop one. 
We can, of course, rescale $\Lambda_{\rm TC}$ freely.  
Here we chose $\Lambda_{\rm TC}$ as 
\begin{equation}
 \Lambda_{\rm TC}:\,\, \alpha(\mu^2=\Lambda_{\rm TC}^2) =
 \frac{\alpha_*}{1+W(e^{-1})} \simeq 0.78\, \alpha_*\, ,
\end{equation}
which reflects the conformal anomaly associated with  
the perturbative running in the UV region $\mu > \Lambda_{\rm TC}$ 
dominated by the one-loop effects, 
\begin{equation}
 \partial_\mu D^\mu\Big|_{\rm perturbative} = 
 \frac{\beta(\alpha)}{4 \alpha^2} \left(\alpha G_{\mu\nu}^2\right) 
 \Big|_{\rm perturbative} = {\cal O} (\Lambda_{\rm TC}^4)\, ,
\label{perturbativeanomaly}
\end{equation}
while keeping the (approximate) conformal symmetry 
(via (almost) nonrunning coupling)  in  the IR region 
$\mu <\Lambda_{\rm TC}$ so as  to be broken only nonperturbatively 
by the dynamical generation of the techni-fermion mass $m$.
Actually, the UV and IR behaviors of $\alpha(\mu^2)$ in 
Eq.~(\ref{alp-lambert}) are
\begin{equation}
  \alpha (\mu^2) \approx \frac{1}{b_0 \ln \frac{\mu^2}{\Lambda_{\rm TC}^2}}
  \qquad (\mu^2 \gg \Lambda_{\rm TC}^2),
\end{equation}
  and
\begin{equation}
  \alpha (\mu^2) \approx
  \frac{\alpha_*}
       {1+e^{-1}\left(\frac{\mu^2}{\Lambda_{\rm TC}^2}\right)^{b_0 \alpha_*}},
  \qquad (\mu^2 \ll \Lambda_{\rm TC}^2) , 
  \label{app-IR}
\end{equation}
respectively. 

\begin{figure}[t]
  \begin{center}
  \resizebox{0.4\textheight}{!}{\includegraphics{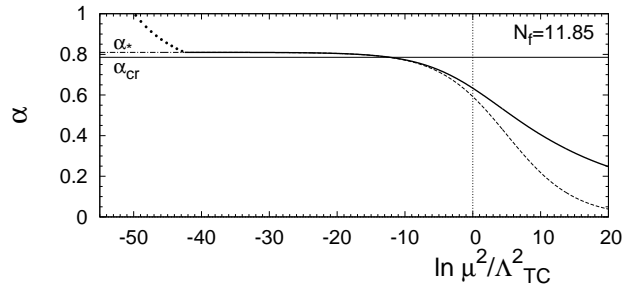}}
  \end{center}
  \caption{Behavior of $\alpha$.
    The bold solid and dashed curves correspond to
    the behavior of $\alpha$ for the two-loop $\beta$ function
    (\ref{banks-zaks}) and the parabolic approximation (\ref{app-parabolic}).
    We took $N_{\rm  TC}=3$ and $N_f=11.85$, which yields $\alpha_*=0.810$.
    The scale $\alpha(\mu^2=\mu_{\rm cr}^2)=\alpha_{\rm cr}=\pi/4$
    is given by $\mu_{\rm cr}/\Lambda_{\rm TC} = 0.00225$.
    Below the scale of the dynamical mass $m$, numerically 
    $m/\Lambda_{\rm TC} = 5.88 \times 10^{-10}$ obtained by solving 
    the corresponding ladder SD equation with 
    the gauge coupling (\ref{alp-lambert}), 
    the techni-fermions should be decoupled and thereby 
    the running of $\alpha$ is expected to be changed.
    The dots below $\mu < m$ corresponds to this expectation. 
    The dash-dotted curve below $\mu < m$ is for the formal 
    solution of the two-loop $\beta$-function.
  \label{fig-alpha}}
\end{figure}

Note that in this paper we are not interested 
in the perturbative part of the conformal anomaly 
in Eq. (\ref{perturbativeanomaly}) and will focus on 
the {\it nonperturbative} contributions to the conformal anomaly 
and the relevant techni-gluon condensate associated 
with the dynamical generation of the mass $m$ in 
the IR dynamics:~\cite{Miransky:1989qc,Haba:2010iv}
\begin{eqnarray}
  \langle\alpha G_{\mu\nu}^2\rangle &\equiv&
  \langle\alpha G_{\mu\nu}^2\rangle_{\rm full}
   - \langle\alpha G_{\mu\nu}^2\rangle_{\rm perturbative} \,,
  \nonumber\\
  \langle \theta_\mu^\mu\rangle &\equiv&
  \langle \theta_\mu^\mu\rangle_{\rm full}
   - \langle \theta_\mu^\mu\rangle_{\rm perturbative} \,,
\label{NPdef}
\end{eqnarray}
where the perturbative conformal anomaly 
$\langle \theta_\mu^\mu\rangle_{\rm perturbative} =
 - {\cal O} (\Lambda_{\rm TC}^4)$ is associated 
with the perturbative running effects of the coupling 
in the UV region $\mu>\Lambda_{\rm TC}$.  
The quantities defined in Eq. (\ref{NPdef})  are similar to 
those discussed in Ref.~\cite{Appelquist:2010gy}.

In order to solve {\it analytically} the improved ladder SD equation,
we would need to simplify the expression of $\alpha$.
We thus adopt the approximation (\ref{app-IR}) in {\it all} region,
because it enjoys both of desirable natures, 
the CBZ-IRFP ($\alpha \to \alpha_*$ for $\mu \to 0$) and
the asymptotic freedom ($\alpha \to 0$ for $\mu \to \infty$).
This approximation corresponds to a parabolic $\beta$ 
function,
\begin{equation}
  \beta(\alpha) = - b_0 \alpha (\alpha_* - \alpha) , 
  \label{app-parabolic}
\end{equation}
which can be applied from 
the IR region to the UV region.
Although the damping of $\alpha$ in the UV region is much faster than 
the two-loop solution (\ref{alp-lambert}), (see Fig.~\ref{fig-alpha},)
it turns out that the critical behavior of the dynamical mass 
is insensitive to the UV behavior of the mass function.
On the other hand, the linear approximation
$\beta (\alpha) = - b_0 \alpha_* (\alpha_* - \alpha)$, 
which yields 
$\alpha(\mu^2) = \alpha_* (1-e^{-1}(\mu^2/\Lambda_{\rm TC}^2)^{b_0\alpha_*})$,
is simpler, but it can be applied only in a narrower region.

Schematic behaviors of the two-loop and parabolic $\beta$ functions
are depicted in Fig.~\ref{fig-beta}.
We also show the running effects of the gauge coupling $\alpha$
for both cases in Fig.~\ref{fig-alpha},
where we took $N_{\rm TC}=3$ and $N_f=11.85$ ($\alpha_*=0.810$).
The parabolic approximation is very successful in the IR region 
$\mu \lesssim \Lambda_{\rm TC}$,
while the damping of $\alpha$ is quicker than the two-loop one
in the UV region $\mu \gg \Lambda_{\rm TC}$,
as we mentioned above. (See Fig.~\ref{fig-alpha}.)
Below the scale of the dynamical mass $m$ discussed in the next section,
the techni-fermions should be decoupled and thereby 
the running of $\alpha$ is expected to be changed.
We depict this expectation by the dots below $\mu < m$ 
in Fig.~\ref{fig-alpha}.

\section{Analysis of the ladder SD equation with running gauge
coupling constants}

\subsection{The CJT potential and the improved ladder SD equation}

\begin{figure}[t]
  \begin{center}
  \resizebox{0.4\textheight}{!}{\includegraphics{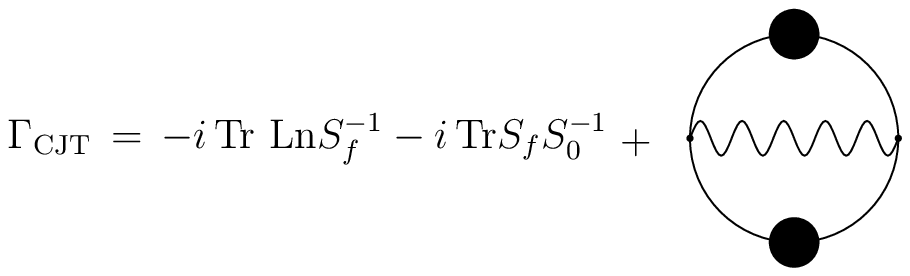}}
  \end{center}
  \caption{Effective action for the fermion propagator.
    The CJT potential is defined by 
    $\Gamma_{\rm CJT}=-V_{\rm CJT}\int dx^4$.
    $S_f$ and $S_0$ represent the full fermion propagator and 
   the free one, respectively.
   In the last diagram, 
   the solid line with a shaded blob and the wavy line represent 
   the full fermion propagator $S_f$ and the gauge boson propagator,
   respectively.
  \label{fig-CJT}}
\end{figure}

The ladder SD approach is a convenient method to analyze 
the dynamical generation of the fermion mass and its critical behavior.
In order to incorporate the running effect of the gauge coupling $\alpha$,
a conventional technique, so-called the improved ladder approximation, 
has been widely employed~\cite{Miransky:1983vj}.
We can derive the improved ladder SD equation via 
the CJT potential $V_{\rm CJT}$~\cite{miransky-textbook}:
\begin{widetext}
\begin{eqnarray}
  V_{\rm CJT}(B) &=& -\frac{N_{\rm  TC} N_f}{4\pi^2}\Bigg[\,
  \int_0^{\Lambda^2} dx x 
  \bigg\{\,\frac{1}{2}\ln \left(1+\frac{B^2(x)}{x}\right)
  - \frac{B^2(x)}{x+B^2(x)}\,\bigg\} \nonumber \\
&& + \frac{1}{2}  \int_0^{\Lambda^2} dx x   \int_0^{\Lambda^2} dy y 
   \frac{B(x) B(y)}{(x+B^2(x))(y+B^2(y))}
   \Bigg(\,\frac{\lambda(x)}{x}\theta (x-y)
  +\frac{\lambda(y)}{y}\theta (y-x)\,\Bigg)\Bigg] , 
   \label{CJT-pot}
\end{eqnarray}
\end{widetext}
with the (normalized) gauge coupling $\lambda(x)$,
\begin{equation}
  \lambda (x) \equiv \frac{3C_F \alpha (\mu^2=x)}{4\pi} ,
\end{equation}
where $x$ and $y$ denote the Euclidean momenta, 
the full fermion propagator inverse is
$iS_f^{-1}(p) = A(-p^2)\fsl{p}-B(-p^2)$,
and we took the Landau gauge at which the fermion wave function
renormalization is unity, $A(-p^2) \equiv 1$. See Fig.~\ref{fig-CJT}.
Although the UV cutoff $\Lambda$ is not needed for 
case of the two-loop running coupling which is asymptotically free in 
the UV region in contrast to the nonrunning case, 
we have put an artificial $\Lambda\, (\rightarrow \infty)$ 
in Eq.(\ref{CJT-pot}) only for the numerical calculation, 
which should not be confused with $\Lambda$ in the nonrunning case 
used in Eqs.(\ref{Miransky}) and (\ref{Miranskybeta}).
The variation of $V_{\rm CJT}$ with respect to 
the fermion mass function $B(x)$ with $x \equiv -p^2$ yields
the improved ladder SD equation~\cite{miransky-textbook},
\begin{equation}
  B(x) = \int_0^{\Lambda^2} dy
  \frac{y B(y)}{y+B^2(y)}\Bigg[\,\frac{\lambda(x)}{x}\theta (x-y)
 +\frac{\lambda(y)}{y}\theta (y-x)\,\Bigg] \, .
  \label{imp-ladder}
\end{equation}

\subsection{Analytic solution for the improved ladder SD equation
in the parabolic approximation}

The integral equation (\ref{imp-ladder}) is equivalent to 
a set of a nonlinear differential equation and boundary conditions (BC's).
It is, however, difficult to solve {\it analytically} the nonlinear 
differential equation in general.
We may adopt the bifurcation method~\cite{Atkinson:1986rm}, 
which yields a more handy linearized differential equation.
We also ignore $xd\lambda/dx \propto \beta$,
because of $\beta \sim 0$ near $x \sim 0$ and $x \sim \infty$.
Under these simplifications, 
we obtain the following differential equation and the two BC's: 
\begin{equation}
     x^2 \frac{d^2}{dx^2} B(x) 
  + 2x  \frac{d}{dx} B(x)
  + \frac{\lambda_*}{1 + e^{-1}\left(\frac{x}{\Lambda_{\rm TC}^2}\right)^s}
    B(x) = 0 , 
 \label{sde-bifurcation}
\end{equation}
and 
\begin{eqnarray}
\mbox{(UV-BC)}: 
&&  x \frac{d}{dx} B(x)\Bigg|_{x=\Lambda^2} + B(\Lambda^2)=0, \\
\mbox{(IR-BC)}: 
&& x^2 \frac{d}{dx} B(x) \Bigg|_{x \to m^2} \to 0 , 
\end{eqnarray}
where $m$ is defined by the normalization of the mass function,
\begin{equation}
  B(x=m^2) = m , 
\end{equation}
and the IRFP $\lambda_*$ and the power factor $s$ are
\begin{equation}
  \lambda_* \equiv \frac{3C_F \alpha_*}{4\pi} , 
\end{equation}
and 
\begin{equation}
  s \equiv b_0 \alpha_* > 0 , 
\end{equation}
respectively.
In the parabolic approximation, 
the normalized gauge coupling $\lambda(x)$ is
\begin{equation}
  \lambda(x) = 
  \frac{\lambda_*}{1 + e^{-1}\left(\frac{x}{\Lambda_{\rm TC}^2}\right)^s} \, .
  \label{lam-parabolic}
\end{equation}
Note that in the limit of $s \to \infty$
the gauge coupling becomes 
$\lambda(x) = \lambda_* \theta (e^{\frac{1}{s}}\Lambda_{\rm TC}^2 - x)$,
where the step function is defined by 
$\theta (t)=1$ for $t > 0$, $\theta (t)=1/2$ for $t = 0$, and
$\theta (t)=0$ for $t < 0$.

We can {\it analytically} solve the differential equation 
(\ref{sde-bifurcation}) as follows:
\begin{widetext}
\begin{eqnarray}
  \frac{B(x)}{m} &=& \phantom{+}
  c_1 \left(\frac{x}{m^2}\right)^{-\frac{1-\omega}{2}} 
  F\bigg(-\frac{1-\omega}{2s},-\frac{-1-\omega}{2s},1+\frac{\omega}{s};
         -\bar{x}^s\bigg)
  \nonumber \\
&&
 +d_1 \left(\frac{x}{m^2}\right)^{-\frac{1+\omega}{2}} 
  F\bigg(-\frac{1+\omega}{2s},-\frac{-1+\omega}{2s},1-\frac{\omega}{s};
         -\bar{x}^s\bigg) , \qquad (x \geq m^2),
  \label{parabolic-sol}
\end{eqnarray}
\end{widetext}
where $F(\alpha,\beta,\gamma;z)$ represents 
the Gauss's hypergeometric function\footnote{
If we employ the linear approximation,
$\lambda(x) = \lambda_* (1-e^{-1}(x/\Lambda_{\rm TC}^2)^s)$,
the analytical solution is written by the modified Bessel functions.}
and we introduced 
\begin{equation}
  \bar{x} \equiv e^{-\frac{1}{s}}\frac{x}{\Lambda_{\rm TC}^2},
\end{equation}
and
\begin{equation}
  \omega \equiv \sqrt{1-\frac{\lambda_*}{\lambda_{\rm cr}}} , \qquad
  \lambda_{\rm cr} \equiv \frac{1}{4} \, .
\end{equation}
The integration constants $c_1$ and $d_1$ are
determined through the IR-BC and the normalization of $B(x)$.
The UV-BC gives the scaling relation.

The normalization $B(x=m^2)=m$ yields
\begin{eqnarray}
1&=& c_1 
  F\bigg(-\frac{1-\omega}{2s},-\frac{-1-\omega}{2s},1+\frac{\omega}{s};
         -\bar{x}_m^s\bigg)
  \nonumber \\
&+& d_1 
  F\bigg(-\frac{1+\omega}{2s},-\frac{-1+\omega}{2s},1-\frac{\omega}{s};
         -\bar{x}_m^s\bigg),
  \label{norm-sol}
\end{eqnarray}
with
\begin{equation}
  \bar{x}_m \equiv e^{-\frac{1}{s}} \frac{m^2}{\Lambda_{\rm TC}^2} \, .
\end{equation}
On the other hand, the IR-BC gives
\begin{eqnarray}
&& \phantom{+}
  c_1 \Bigg[\,\omega
  F\bigg(-\frac{1-\omega}{2s},-\frac{-1-\omega}{2s},1+\frac{\omega}{s};
         -\bar{x}_m^s\bigg)
  \nonumber \\
&&
  +\frac{\lambda_*}{s+\omega} \bar{x}_m^s
   F\bigg(1-\frac{1-\omega}{2s},1+\frac{1+\omega}{2s},2+\frac{\omega}{s};
         -\bar{x}_m^s\bigg)\,\Bigg]
  \nonumber \\
&&
 +d_1 \frac{\lambda_*}{s-\omega} \bar{x}_m^s
   F\bigg(1-\frac{1+\omega}{2s},1+\frac{1-\omega}{2s},2-\frac{\omega}{s};
         -\bar{x}_m^s\bigg)
  \nonumber \\
&& = \frac{1+\omega}{2},
\end{eqnarray}
where we used Eq.~(\ref{norm-sol}).

In the limit of $m \ll \Lambda_{\rm TC}$, we obtain
\begin{equation}
  c_1 = \frac{1+\omega}{2\omega}, \qquad
  d_1 = -\frac{1-\omega}{2\omega} ,   
\end{equation}
which corresponds to the coefficients of the bifurcation solution
with the fixed gauge coupling $\lambda(x) = \lambda_*$.
On the other hand, the UV-BC in the limit of 
$m^2 \ll \Lambda_{\rm TC}^2 \ll \Lambda^2$ yields
\begin{equation}
  \left(\frac{e^{\frac{1}{s}}\Lambda_{\rm TC}^2}{m^2}\right)^{\omega}
 =\frac{(1-\omega)^2}{(1+\omega)^2}
  \frac{\Gamma\bigg(1-\frac{\omega}{s}\bigg)
        \Gamma^2\bigg(1+\frac{1+\omega}{2s}\bigg)}
       {\Gamma\bigg(1+\frac{\omega}{s}\bigg)
        \Gamma^2\bigg(1+\frac{1-\omega}{2s}\bigg)} \, .
  \label{UV-BC-app1}
\end{equation}
It is noticeable that the dependence of the UV cutoff $\Lambda$ 
disappears.
Only when $\omega$ is pure imaginary, i.e.,
\begin{equation}
  \lambda_* > \lambda_{\rm cr} = \frac{1}{4},
\end{equation}
Eq.~(\ref{UV-BC-app1}) has a relevant solution,
\begin{eqnarray}
  \ln \frac{m}{e^{\frac{1}{2s}}\Lambda_{\rm TC}}
& = &
   - \frac{n\pi}{\tilde{\omega}}
   + \frac{2\arctan \tilde{\omega}}{\tilde{\omega}}
     \nonumber \\
&& - \frac{\ln \Bigg[\,
                 \frac{\Gamma\bigg(1-\frac{i\tilde{\omega}}{s}\bigg)
                 \Gamma^2\bigg(1+\frac{1+i\tilde{\omega}}{2s}\bigg)}
                {\Gamma\bigg(1+\frac{i\tilde{\omega}}{s}\bigg)
                 \Gamma^2\bigg(1+\frac{1-i\tilde{\omega}}{2s}\bigg)}\,\Bigg]}
     {2i\tilde{\omega}} , 
  \label{scaling-app1}
\end{eqnarray}
where 
\begin{equation}
  \tilde{\omega} \equiv \sqrt{\frac{\lambda_*}{\lambda_{\rm cr}} - 1} , 
\end{equation}
and $n=1,2,3,\cdots$.
It is known that the zero node solution $n=1$ is 
the true vacuum~\cite{miransky-textbook}.
Eq.~(\ref{scaling-app1}) yields 
the essential singularity scaling relation, 
\begin{equation}
 m \sim  e^{-\frac{\pi}{\tilde{\omega}}} \,\Lambda_{\rm TC}=\Lambda_{\rm TC} \cdot \exp \left(-\frac{\pi}{
 \sqrt{\frac{\alpha_*}{\alpha_{\rm cr}}-1}}\right)\, , 
 \label{ess-sing}
\end{equation}
similarly to Eq.~(\ref{Miransky}) with replacement of $\alpha$ by $\alpha_*$ and $\Lambda$ by $\Lambda_{\rm TC}$ .

The behavior of the mass function $B(x)$ in 
the supercritical region $\lambda_* > \lambda_{\rm cr}$
is approximately
\begin{eqnarray}
  \frac{B(x \ll \Lambda_{\rm TC}^2)}{m} & \simeq & 
  \frac{\sqrt{1+\tilde{\omega}^2}}{\tilde{\omega}}
  \left(\frac{x}{m^2}\right)^{-\frac{1}{2}}
  \nonumber \\ &&
  \sin \Bigg[\,\frac{\tilde{\omega}}{2} \ln \left(\frac{x}{m^2}\right)
 + \arctan \tilde{\omega}\,\Bigg], 
 \label{Bsol-IR} \\
  \frac{B(x \gg \Lambda_{\rm TC}^2)}{m} & \simeq & 
   e^{\frac{1}{2s}} 
   \Bigg[\,A(\tilde{\omega})  + A(-\tilde{\omega})\,\Bigg]
   \frac{m \Lambda_{\rm TC}} {x} ,
  \label{Bsol-UV} 
\end{eqnarray}
with
\begin{equation}
  A(\tilde{\omega}) \equiv
  \frac{\lambda_*}{i\tilde{\omega}}
      \frac{\Gamma\bigg(1+\frac{i\tilde{\omega}}{s}\bigg)
            \Gamma\bigg(1-\frac{1}{s}\bigg)}
           {\Gamma^2\bigg(1+\frac{-1+i\tilde{\omega}}{2s}\bigg)}
     \left(\frac{e^{\frac{1}{2s}}\Lambda_{\rm TC}}{m}\right)^{i\tilde{\omega}} \, .
\end{equation}
The behaviors of the mass function in the IR and UV regions
correspond to those with the anomalous dimensions 
$\gamma_m=1$ and $\gamma_m=0$, respectively.
In particular, owing to the quicker damping of $\alpha$ than 
the logarithm, there is no log correction unlike QCD.
On the other hand, the IR behavior is the same as that for 
the fixed coupling.

In passing, the critical number $N_f$, which corresponds to 
$\lambda_{\rm cr}$, is
\begin{equation}
  N_f^{\rm cr} = 4N_{\rm TC}\Bigg[\,1-\frac{3}{10}\frac{1}{5N_{\rm TC}^2-3}\,\Bigg] \, .
\end{equation}
Since the power factor $s$ is
\begin{equation}
  s = b_0 \alpha_* = \frac{(11N_{\rm TC}-2N_f)^2}{-6[17N_{\rm TC}^2-N_f(5N_{\rm TC}+3C_F)]},
\end{equation}
at the critical point, it reads
\begin{equation}
  s_{\rm cr} = b_0 \alpha_{\rm cr}  
 = \frac{N_{\rm TC}}{18(N_{\rm  TC}^2-1)}
   \Bigg[\,3N_{\rm  TC} + \frac{12}{5}\frac{N_{\rm TC}}{5N_{\rm TC}^2-3}\,\Bigg] \, .
\end{equation}
For $N_{\rm TC}=3$, they are numerically
\begin{equation}
  N_f^{\rm cr} = \frac{417}{35} \simeq 11.914, \qquad
  s_{\rm cr} = \frac{107}{560} \simeq 0.19102 \, .
\end{equation}

We can solve numerically the improved ladder SD equation 
(\ref{imp-ladder}) with the normalized gauge coupling (\ref{lam-parabolic}).
The computational technique is described in Ref.~\cite{Hashimoto:2000uk}.

\begin{figure}[t]
  \begin{center}
  \resizebox{0.4\textheight}{!}{\includegraphics{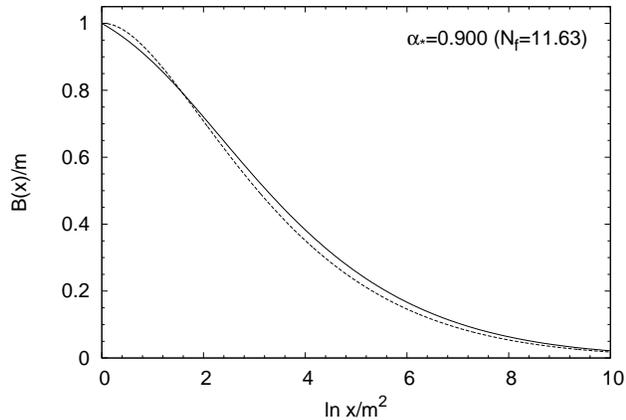}}
  \end{center}
  \caption{Behaviors of the mass function in the parabolic approximation.
    The solid and dashed curves correspond to
    the numerical solution of the improved ladder SD equation 
    (\ref{imp-ladder}) with the running gauge coupling (\ref{lam-parabolic}) 
    and the bifurcation solution (\ref{parabolic-sol}), 
    respectively.
    We took $N_{\rm TC}=3$ and $N_f=11.63$, which yields
    $\alpha_*=0.900$ and $\lambda_*=0.287$.
  \label{fig-massfunc-parabilic}}
\end{figure}

We depict the analytical and numerical solutions of $B(x)$ in 
Fig.~\ref{fig-massfunc-parabilic}, where we took $N_{\rm TC}=3$ and $N_f=11.63$.
Although we drastically simplified the integral equation (\ref{imp-ladder})
into the linearized differential equation (\ref{sde-bifurcation})
with the two BC's, we find that the approximation works well.

\begin{figure}[t]
  \begin{center}
  \resizebox{0.4\textheight}{!}{\includegraphics{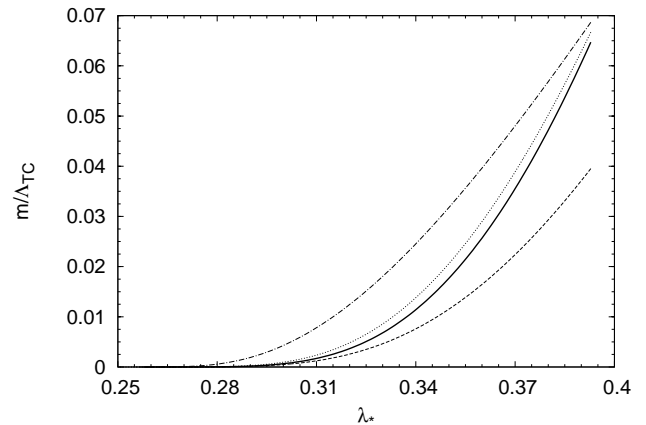}}
  \end{center}
  \caption{Scaling relations in several approaches.
    The solid, dashed and dash-dotted curves correspond to
    the numerical solution of the ladder SD equation (\ref{imp-ladder}) for
    the two-loop $\beta$ function, the parabolic approximation, and
    the fixed coupling, respectively.
    The dotted one is for the analytical expression (\ref{scaling-app1})
    in the parabolic approximation.
    We took $N_{\rm  TC}=3$ and varied continuously $N_f$.
  \label{fig-scaling-BK-para-FP}}
\end{figure}

The scaling relations in the numerical and analytical 
approaches are shown as the dashed and dotted curves 
in Fig.~\ref{fig-scaling-BK-para-FP}, respectively.
We confirmed that the numerical solution is unchanged for
$\Lambda/\Lambda_{\rm TC} = 10^{1,2,\cdots,10}$.
(It is not the case for $\Lambda=\Lambda_{\rm TC}$, however.)
In the figure, we took $\Lambda/\Lambda_{\rm TC} = 10^{5}$.
The shapes of the scaling relation are qualitatively similar. 
We also find that the analytical expression (\ref{scaling-app1})
is close to the numerical solution for 
the two-loop gauge coupling (\ref{alp-lambert}),
which will be discussed in the next subsection.

\subsection{Numerical solution for the improved ladder SD equation 
with the two-loop gauge coupling}

We studied the parabolic approximation in the analytical and 
numerical ways, so far.

\begin{figure}[t]
  \begin{center}
  \resizebox{0.4\textheight}{!}{\includegraphics{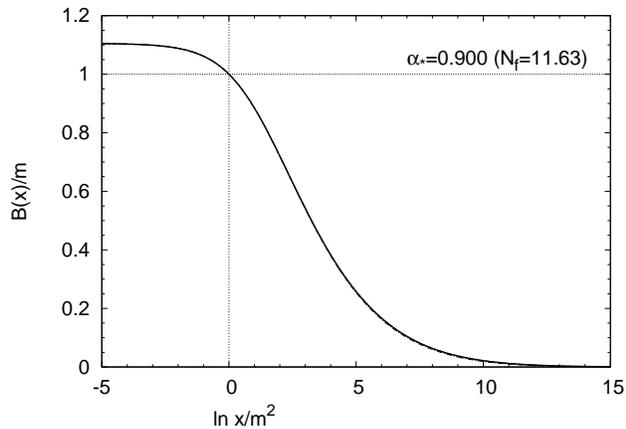}}
  \end{center}
  \caption{Behaviors of the mass function in the numerical approaches.
    The bold solid, dashed and dash-dotted curves correspond to
    the numerical solutions of the improved ladder SD equation for
    the two-loop $\beta$ function, the parabolic approximation, and
    the fixed coupling, respectively.
    We normalized each scale by each dynamical mass $B(x=m^2)=m$.
    We took $N_{\rm TC}=3$ and $N_f=11.63$, which yields
    $\alpha_*=0.900$ and $\lambda_*=0.287$.
    In this resolution, we cannot distinguish each other, however.
  \label{fig-massfunc-BK-para-FP}}
\end{figure}

\begin{figure}[t]
  \begin{center}
  \resizebox{0.4\textheight}{!}{\includegraphics{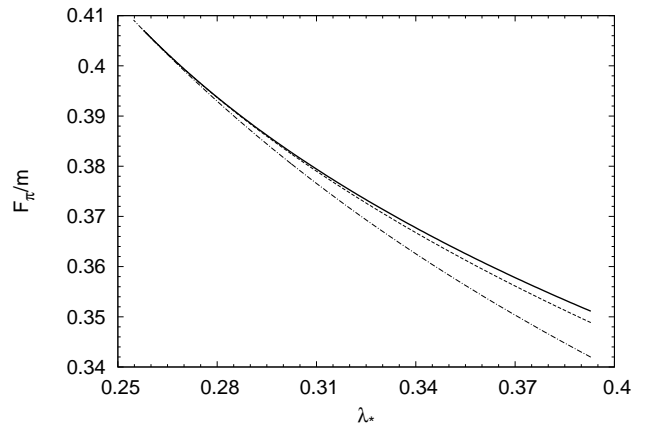}}
  \end{center}
  \caption{Decay constant $F_\pi$ in the numerical approaches.
    The solid, dashed and dash-dotted curves correspond to
    the numerical solution of the ladder SD equation (\ref{imp-ladder}) for
    the two-loop $\beta$ function, the parabolic approximation, and
    the fixed coupling, respectively.
    We took $N_{\rm  TC}=3$ and $N_D=1$.
  \label{fig-fpi-BK-para-FP}}
\end{figure}

\begin{figure}[t]
  \begin{center}
  \resizebox{0.25\textheight}{!}{\includegraphics{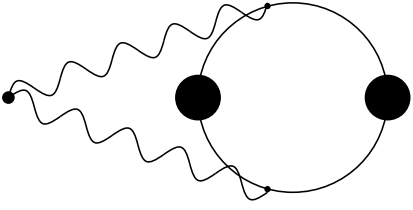}}
  \end{center}
  \caption{Techni-gluon condensate associated with  the generation of mass $m$.
  \label{fig-glu-cond}}
\end{figure}

\begin{figure}[t]
  \begin{center}
  \resizebox{0.4\textheight}{!}{\includegraphics{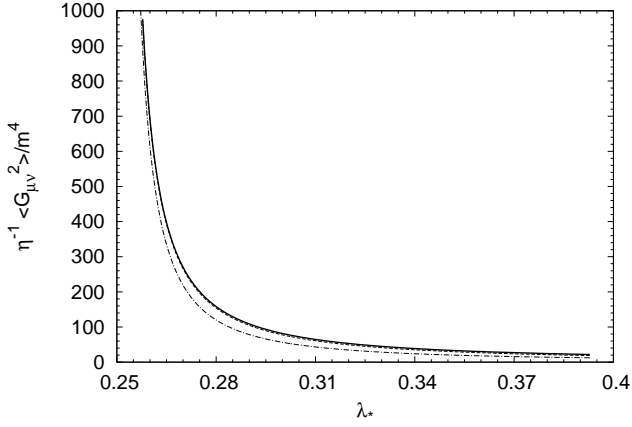}}
  \end{center}
  \caption{Behaviors of the gluon condensate in the numerical approaches.
    The solid, dashed and dash-dotted curves correspond to
    the numerical solution of the ladder SD equation (\ref{imp-ladder}) for
    the two-loop $\beta$ function, the parabolic approximation, and
    the fixed coupling, respectively.
    The factor $\eta$ is defined by $\eta \equiv N_{\rm TC} N_f/(2\pi^2)$.
    We took $N_{\rm TC}=3$.
    The cutoff is $\Lambda/\Lambda_{\rm TC}=10^{5}$ for
    the two-loop $\beta$ function and the parabolic approximation.
    For the fixed gauge coupling, the cutoff is taken as
    $\Lambda=\Lambda_{\rm TC}$.
  \label{fig-G2-BK-para-FP}}
\end{figure}

\begin{figure}[t]
  \begin{center}
  \resizebox{0.4\textheight}{!}{\includegraphics{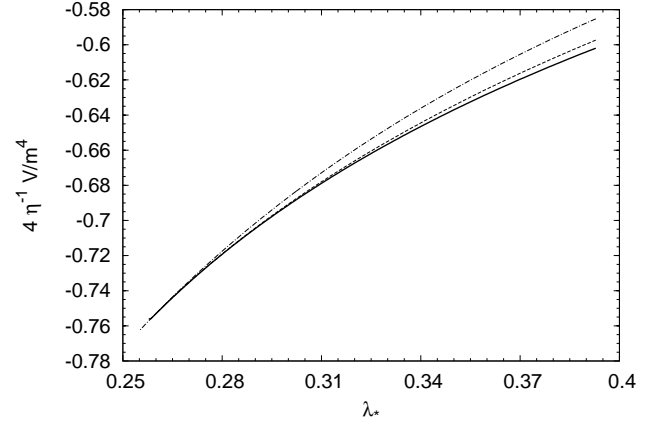}}
  \end{center}
  \caption{Behaviors of the vacuum energy in the numerical approaches.
    The solid, dashed and dash-dotted curves correspond to
    the numerical solutions of the ladder SD equation (\ref{imp-ladder}) for
    the two-loop $\beta$ function, the parabolic approximation, and
    the fixed coupling, respectively.
    The factor $\eta$ is defined by $\eta \equiv N_{\rm TC} N_f/(2\pi^2)$.
    We took $N_{\rm  TC}=3$.
  \label{fig-vac-BK-para-FP}}
\end{figure}

\begin{figure}[t]
  \begin{center}
  \resizebox{0.4\textheight}{!}{\includegraphics{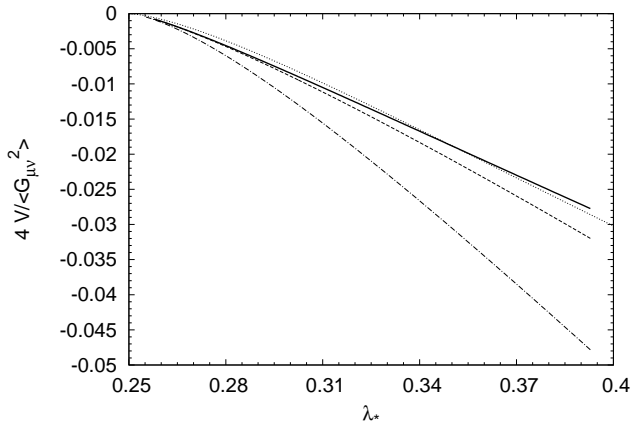}}
  \end{center}
  \caption{Ratio of the vacuum energy and the techni-gluon condensate.
    The solid, dashed and dash-dotted curves correspond to
    the numerical solutions of the ladder SD equation (\ref{imp-ladder}) for
    the two-loop $\beta$ function, the parabolic approximation, and
    the fixed coupling, respectively.
    We took $N_{\rm  TC}=3$.
    The dotted line corresponds to the least-squares fitting by
    $\xi \tilde{\omega}^3/\lambda_*$.
    Numerically, we obtain $\xi \simeq -0.026$.
  \label{fig-vac-G2-BK-para-FP}}
\end{figure}

\begin{figure}[t]
  \begin{center}
  \resizebox{0.4\textheight}{!}{\includegraphics{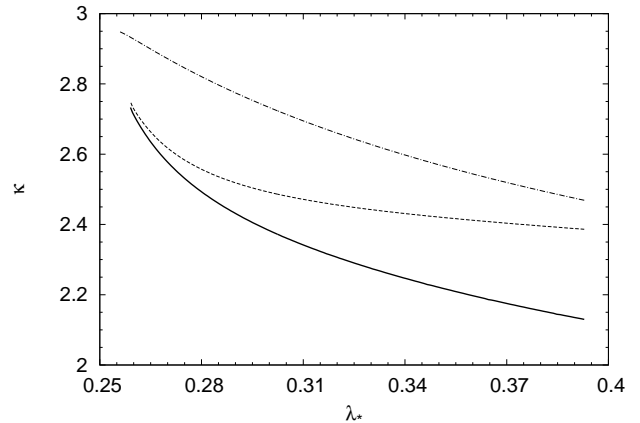}}
  \end{center}
  \caption{Instantaneous exponent of the nonperturbative beta function 
    with respect to $\tilde{\omega}$. 
    The vertical axis at $\lambda_*=1/4=0.25$ is the criticality
    (conformal edge).
    The solid, dashed and dash-dotted curves correspond to
    the numerical solutions of the ladder SD equation (\ref{imp-ladder}) for
    the two-loop $\beta$ function, the parabolic approximation, and
    the fixed coupling, respectively.
  \label{fig-kappa}}
\end{figure}

Let us now solve numerically the improved ladder SD equation with 
the gauge coupling (\ref{alp-lambert}) expressed through the Lambert function.
We calculate the Lambert function via 
the Halley's method~\cite{Corless:1996zz},
\begin{equation}
  w_{j+1}=w_j
  - \frac{w_j e^{w_j}-z}{e^{w_j}(w_j+1)-\frac{(w_j+2)(w_j e^{w_j}-z)}{2w_j+2}}\,.
\end{equation}
The computational technique for solving the improved ladder SD equation
is described in Ref.~\cite{Hashimoto:2000uk}.
The results are depicted in 
Figs.~\ref{fig-scaling-BK-para-FP}--\ref{fig-kappa}.
We also show the results for the fixed gauge coupling $\lambda(x)=\lambda_*$.
In this case, we take the UV cutoff $\Lambda$ of the SD equation
to $\Lambda_{\rm TC}$.
We confirmed that the consequences of the fixed gauge coupling
are consistent with those in Ref.~\cite{Harada:2003dc}, 
where $\lambda(x)=\lambda_* \theta(\Lambda_{\rm TC}^2-x)$ was 
essentially used, instead of the two-loop one (\ref{alp-lambert}).

We depict the scaling relation in Fig.~\ref{fig-scaling-BK-para-FP}.
We confirmed that the numerical solution for the two-loop gauge coupling
(\ref{alp-lambert}) is unchanged for $\Lambda/\Lambda_{\rm TC} = 10^{2,\cdots,10}$.
(It is not the case for $\Lambda/\Lambda_{\rm TC} = 10^{0,1}$, however.)
In the figure, we took $\Lambda/\Lambda_{\rm TC} = 10^{5}$.
We find that the numerical values of the dynamical mass $m/\Lambda_{\rm TC}$ 
for the two-loop gauge coupling is smaller than 
those for the fixed coupling.
It is amazing that the analytic solution for the parabolic approximation
is quantitatively close to the numerical one for the two-loop gauge coupling.

In Fig.~\ref{fig-massfunc-BK-para-FP},
we show the behaviors of the mass functions for the two-loop gauge coupling,
the parabolic approximation, and the fixed gauge coupling.
In this resolution, we cannot distinguish each other.
We did not draw here the analytical solution (\ref{parabolic-sol}) 
for the parabolic approximation.
Although the behavior is close to the numerical one,
there is a slight deviation between the analytical 
and three numerical solutions. 
Compare Fig.~\ref{fig-massfunc-parabilic} with 
Fig.~\ref{fig-massfunc-BK-para-FP}.
An important point is that we normalized the mass function by 
the dynamical mass $B(x=m^2)=m$, not by $\Lambda_{\rm TC}$.
Note that the dynamical masses for the two-loop gauge coupling, 
the parabolic approximation, 
and the fixed gauge coupling are numerically obtained as
$m/\Lambda_{\rm TC} = 1.08 \times 10^{-4}, 0.845 \times 10^{-4}, 
 13.3 \times 10^{-4}$ for $N_{\rm TC}=3$ and $N_f=11.63$, respectively.
If we had normalized $B(x)$ by $\Lambda_{\rm TC}$, 
the three behaviors would thus look very different.
Owing to this universal nature of the dimensionless mass function 
normalized by the dynamical mass, $B(x)/m$, 
the normalized physical quantities such as the decay constant $F_\pi/m$, 
the vacuum energy $\VEV{\theta_\mu^\mu}/m^4$, and
the techni-gluon condensate $\VEV{G_{\mu\nu}^2}/m^4$, which 
are determined through the mass function,
become insensitive to the approximations of the running gauge coupling
near the conformal edge, as we will see later.

How about the relation between the scale 
$\alpha(\mu^2=\mu_{\rm cr}^2) = \alpha_{\rm cr}$ and 
the dynamical mass $m$? 
By definition, we can obtain the scale $\mu_{\rm cr}$ by
\begin{equation}
  W(z_{\rm cr}) = 4\lambda_* - 1,
\end{equation}
with
\begin{equation}
  z_{\rm cr} \equiv 
 e^{-1} \left(\frac{\mu_{\rm cr}^2}{\Lambda_{\rm TC}^2}\right)^{b_0\alpha_*} \, .
\end{equation}
For $N_{\rm TC}=3$ and $N_f=11.63$, 
the scale $\alpha(\mu^2=\mu_{\rm cr}^2)=\alpha_{\rm cr}$
is numerically obtained as $\mu_{\rm cr}/\Lambda_{\rm TC} = 0.189$.
As we showed previously, the dynamical mass is 
$m/\Lambda_{\rm TC} \simeq 1.08 \times 10^{-4}$.
When we vary the number of flavor to $N_f=11.85$,
they are much more hierarchical, 
$\mu_{\rm cr}/\Lambda_{\rm TC} = 0.00225$ and 
$m/\Lambda_{\rm TC} \simeq 5.88 \times 10^{-10}$.
In the parabolic approximation, we find a more handy formula,
\begin{equation}
  \mu_{\rm cr} = e^{\frac{1}{2s} (\ln \tilde{\omega}+1)} \Lambda_{\rm TC} \, .
\end{equation}
This is apparently much larger than 
the dynamical mass $m$ in Eq.(\ref{ess-sing}),
\begin{equation}
    m \sim e^{-\frac{\pi}{\tilde{\omega}}} \Lambda_{\rm TC} \, .
\end{equation}
We show a concrete value for $N_{\rm TC}=3$ and $N_f=11.85$
in Fig.~\ref{fig-alpha}.

Let us calculate the decay constant $F_\pi$ of the techni-pion, which is 
connected with the weak boson mass.
We assume that a part of the fermion flavor $N_f$ couples to 
the electroweak current.
In order to estimate the decay constant, we employ
the Pagels-Stokar formula~\cite{Pagels:1979hd},
\begin{equation}
   F_\pi^2 = \frac{N_D N_{\rm TC}}{4\pi^2}
  \int_{0}^{\Lambda^2}dx x
  \frac{B^2(x)-\frac{x}{4}\frac{d}{dx}B^2(x)}
       {[x+B^2(x)]^2}, 
       \label{PS}
\end{equation}
where $N_D$ denotes the number of fermion doublets which couple to 
the electroweak current.
The numerical results for the two-loop gauge coupling, 
the parabolic approximation, 
and the fixed gauge coupling are shown in Fig.~\ref{fig-fpi-BK-para-FP}.
We found that the parabolic approximation works well. 
Note that $F_\pi/m \simeq 0.41 \times \sqrt{N_D N_{\rm TC}/3}$ 
near the critical coupling, where we took into account 
the dependence of $N_{\rm TC}$ and $N_D$.
Thus, when we fix $F_\pi = 246$~GeV, we can estimate the dynamical mass
as $m \sim \mbox{1 TeV}/\sqrt{N_D N_{\rm  TC}}$.

The non-perturbative techni-gluon condensate defined in Eq.(\ref{NPdef}) 
can be estimated as~\cite{Gusynin:1988cf}
\begin{equation}
 \VEV{G_{\mu\nu}^2} 
  = \frac{N_{\rm TC} N_f}{2\pi^2}
  \int_{0}^{\Lambda^2}dx x \frac{B^2(x)}{x+B^2(x)} \, .
  \label{glu-cond}
\end{equation}
See also Fig.~\ref{fig-glu-cond}.
Note that after substituting the solution of the ladder SD equation,
the expression of  $\VEV{G_{\mu\nu}^2}$ explicitly depends 
only on the mass function $B(x)$ and has no explicit dependence of 
the running of the gauge coupling~\cite{Gusynin:1988cf}.
Hence we expect that the result is not sensitive to the details of 
the running behaviors of the gauge coupling: 
The numerical results for the two-loop gauge coupling, 
the parabolic approximation, 
and the fixed gauge coupling are shown in Fig.~\ref{fig-G2-BK-para-FP}.
We found that the behavior of  $\VEV{G_{\mu\nu}^2}$ is {\it not} like
$\VEV{G_{\mu\nu}^2} \sim m^4$ as assumed in 
Refs.~\cite{Bando:1986bg, Appelquist:2010gy},
but $\VEV{G_{\mu\nu}^2}/m^4 \sim 1/\tilde{\omega}^3 \to \infty$ 
in the limit of $\lambda_* \to \lambda_{\rm cr}$. 
Our result directly confirms the estimate of the techni-gluon 
condensate made in Ref.~\cite{Haba:2010iv}
which assumed the ladder result for the vacuum energy 
Eq.(\ref{vacuumenergy}) and nonperturbative beta function 
Eq.(\ref{Miranskybeta}) in the case of  the nonrunning coupling.
We can show this behavior by using the approximation (\ref{Bsol-IR}), 
i.e.,
\begin{equation}
  \VEV{G_{\mu\nu}^2} 
  \simeq \frac{N_{\rm TC} N_f}{2\pi^2}\int_{m^2}^{\Lambda_{\rm TC}^2} dx B^2(x),
\end{equation}
and thus
\begin{equation}
  \VEV{G_{\mu\nu}^2}
  \simeq \frac{N_{\rm TC} N_f}{2\pi^2}
  \frac{1+\tilde{\omega}^2}{\tilde{\omega}^2} m^4
  \ln \left(\frac{\Lambda_{\rm TC}}{m}\right) 
  \sim \frac{N_{\rm TC} N_f}{2\pi} \frac{m^4}{\tilde{\omega}^3} , 
\end{equation}
where we used the scaling relation (\ref{ess-sing}).

Next to the vacuum energy $V =\langle \theta_\mu^\mu\rangle /4$ 
as defined in Eq.(\ref{NPdef}).
Substituting the solution $B_{\rm sol}(x)$ of 
the ladder SD equation (\ref{imp-ladder}) 
for the CJT potential (\ref{CJT-pot}),
we obtain the vacuum energy,
\begin{eqnarray}
 V &=& V_{\rm CJT}(B=B_{\rm sol}) , \\
   &=& -\frac{N_{\rm  TC} N_f}{8\pi^2}
        \int_0^{\Lambda^2} dx x 
        \Bigg[\,\ln \left(1+\frac{B_{\rm sol}^2(x)}{x}\right)
   \nonumber \\
&& \hspace*{3.5cm}
  - \frac{B_{\rm sol}^2(x)}{x+B_{\rm sol}^2(x)}\,\Bigg] ,  
  \label{vac-energy}
\end{eqnarray}
where we explicitly wrote the subscript for the mass function
in order to distinguish the vacuum energy from the CJT potential itself.
(Of course, also in the expressions (\ref{PS}) and (\ref{glu-cond}),
$B(x)$ represents $B(x)=B_{\rm sol}(x)$.)
The numerical results for the two-loop gauge coupling, 
the parabolic approximation, 
and the fixed gauge coupling are shown in Fig.~\ref{fig-vac-BK-para-FP}.
It is clear that the vacuum energy normalized by $m^4$ does not vanish,
because the numerical calculations shown in Fig.~\ref{fig-vac-BK-para-FP}
suggests
\begin{equation}
 \langle \theta_\mu^\mu\rangle=  4 V \simeq - 0.76 \, \eta\, m^4, 
 \quad \mbox{with} \quad
 \eta \equiv \frac{N_{\rm TC} N_f}{2\pi^2},
 \label{vac-num-2loop}
\end{equation}
near the critical coupling.
This result disagrees with the assumption 
in Refs.~\cite{Bando:1986bg, Appelquist:2010gy}.

The approximate expression (\ref{Bsol-IR}) suggests that 
our results coincide with the estimate of $V$ for 
the fixed gauge coupling~\cite{Miransky:1989qc}:
\begin{equation}
  4 V \simeq - \frac{4 N_{\rm TC}N_f}{\pi^4} m^4  = - 0.81\, \eta\, m^4 \, .
   \label{IR-app}
\end{equation}
It is to be noted that this value is also close to 
the numerical estimate (\ref{vac-num-2loop}) in our case.  In fact,
although the CJT potential itself explicitly depends on
the running of the gauge coupling,
the vacuum energy has an explicit dependence only on 
the mass function $B_{\rm sol}(x)$
after using the solution of the SD equation,
and hence only depends implicitly on the running gauge coupling 
through $B_{\rm sol}(x)$.
(Compare Eq.~(\ref{CJT-pot}) with Eq.~(\ref{vac-energy}).)
The IR behaviors of $B_{\rm sol}(x)$ 
for the two-loop running and fixed couplings coincide
each other, as shown in Fig.~\ref{fig-massfunc-BK-para-FP}. 
(Inside of the frame of the figure corresponds to the IR region.)
Since the UV contribution to the vacuum energy is negligible,
the vacuum energy (\ref{vac-num-2loop}) for the two-loop running 
gauge coupling is almost the same as that for the nonrunning one.

{}From these analytical and numerical analyses,
we conclude that in the vacuum energy there is no divergence
unlike the techni-gluon condensate, $\VEV{G_{\mu\nu}^2}/m^4 \to \infty$,
and also the quantity $V/m^4$ does not approach to zero in the limit of 
$\lambda_* \to \lambda_{\rm cr}$, within our approximation.

Since the formal RG analysis yields~\cite{Miransky:1996pd}
\begin{equation}
 \VEV{\theta_\mu^\mu} = 4 V = \frac{\beta}{4\alpha}\VEV{G_{\mu\nu}^2} , 
  \label{vac-G2}
\end{equation}
the ratio $4 V/\VEV{G_{\mu\nu}^2}$ is closely connected with 
the $\beta$ function.
We depict it in Fig.~\ref{fig-vac-G2-BK-para-FP}.
By using the least-squares method, 
we numerically obtain 
\begin{equation}
  \frac{\beta}{4\alpha} =  \frac{4 V}{\VEV{G_{\mu\nu}^2}} = 
  \xi \frac{\tilde{\omega}^3}{\lambda_*} , \quad \mbox{with} \quad
  \xi = -0.026 \, .
  \label{NP-beta-2loop}
\end{equation}
in the form similar to the nonrunning case. In  the case of 
the nonrunning coupling $\lambda(x)=\lambda_*$,  
we have the analytical result $\xi = -1/(8\pi) = -0.0398$
corresponding to Eq.(\ref{Miranskybeta}), 
$\beta = - 2\tilde{\omega}^3/(3C_F)$, which agrees with 
the numerical result  $\xi = -0.0400$ in the vicinity of 
the critical coupling. 
Incidentally, in the case of the two-loop coupling 
we may take a fitting function other than the above:   
$\xi (1/\lambda_{\rm cr}-1/\lambda_*)^{3/2}$ with $\xi=-0.0155$, which  yields
much better fitting. 

How about the exponent of the nonperturbative beta function
with respect to $\tilde{\omega}$?
Let us write 
\begin{equation}
  \frac{\beta}{4\alpha} =  \frac{4 V}{\VEV{G_{\mu\nu}^2}} = 
  f(\lambda_*) \, \tilde{\omega}^\kappa , 
\end{equation}
where $f(\lambda_*)$ represents some function of $\lambda_*$ and
the instantaneous exponent $\kappa=\kappa(\lambda_*)$ is 
extracted from the relation
\begin{equation}
 \kappa=  \frac{\tilde{\omega}\frac{\partial }{\partial \tilde{\omega}}
        \left(\frac{\beta}{4\alpha}\right)}
       {\left(\frac{\beta}{4\alpha}\right)} 
       \, .
\end{equation}
We depict the numerical results in Fig.~\ref{fig-kappa}.
Near the critical edge, the numerical values of $\kappa=\kappa(\lambda_*)$ are
\begin{equation}
  \kappa \simeq 2.73, \; 2.75, \; 2.95,
\end{equation}
for the two-loop exact solution, the parabolic approximation
and the fixed coupling case, respectively.
Because the linear and multiple zeros of 
the beta function with respect to 
$\alpha(\mu) = \alpha \simeq \alpha_*$ 
correspond to $\kappa|_{\lambda_*=\lambda_{\rm cr}} = 2$ and 
$\kappa|_{\lambda_*=\lambda_{\rm cr}}  > 2$, respectively,
the numerical results obviously show that 
the nonperturbative beta function has the multiple zero 
at the critical edge.
If we smoothly extrapolate the behavior of $\kappa$ to the criticality,
the behavior of the nonperturbative beta function at the criticality 
will be $\beta \propto - \tilde{\omega}^3$,
in accord with the above least-square fitting (\ref{NP-beta-2loop}).

\begin{figure}[t]
  \begin{center}
  \resizebox{0.4\textheight}{!}{\includegraphics{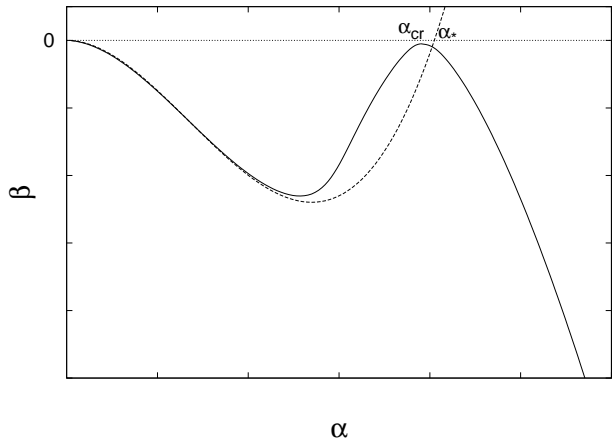}}
  \end{center}
  \caption{Conjecture of the shape of the $\beta$ function
  including the nonperturbative behavior. 
  The bold solid and dashed curves correspond to 
  the conjectured $\beta$ function and the purely perturbative one, 
  respectively.
  \label{fig-beta-conj}}
\end{figure}

It is noticeable that the nonperturbative beta function has 
a multiple zero at the critical coupling 
$\alpha \simeq \alpha_* = \alpha_{\rm cr}$, 
as shown in Eq.~(\ref{NP-beta-2loop}),
which corresponds to the essential singularity scaling 
Eq.~(\ref{ess-sing}).
On the other hand, it is not the case in 
the perturbative (two-loop) expression (\ref{banks-zaks}),
which has a linear zero, 
$\beta \sim \alpha - \alpha_* \sim \alpha_{\rm cr}-\alpha_*$
at criticality $\alpha = \alpha_{\rm cr}$.
Therefore the actual beta function is crucially 
different from the perturbative beta function 
which should be modified in the IR region 
where the non-perturbative dynamics responsible for  
the mass generation is dominant. 
The full $\beta$ function including the perturbative and 
the nonperturbative contributions is thus suggested in 
Fig.~\ref{fig-beta-conj}.
We hope that the lattice studies will clarify this nature.

Now we discuss the behavior of the TD mass in the limit toward the criticality.
Through the PCDC, Eq.(\ref{PCDC}), 
the vacuum energy is connected with the TD mass,
\begin{equation}
  F_{\rm TD}^2 M_{\rm TD}^2 = -d_\theta \VEV{\theta_\mu^\mu} =-4 d_\theta V,
  \label{md}
\end{equation}
where $F_{\rm TD}$, $M_{\rm TD}$ and $d_\theta (=4)$ represent
the TD decay constant, the TD mass and 
the scale dimension of the trace of the energy-momentum tensor. 
Our results qualitatively agree with  the 
conclusion in the case of  SD equation with the nonrunning gauge coupling,
Eq.(\ref{PCDC2}), which is in disagreement with 
Refs.~\cite{Bando:1986bg, Appelquist:2010gy}:
There is no true (massless) NG boson for the conformal symmetry 
at the criticality, unless the TD decay constant diverges, 
i.e., gets decoupled. 
Such a decoupled TD was in fact implied by the idealized limit of 
the holographic TD~\cite{Haba:2010iv}.
In the realistic situation of the TC model building $m/\Lambda_{\rm TC}$ 
is not arbitrarily small but only 
$m/\Lambda_{\rm TC}\sim m/\Lambda_{\rm ETC} \sim 10^{-3}-10^{-4}$,
so that the ``masslessness'' and ``decoupling'' are somewhat milder.
In the quantitative sense our results, though valid only in 
the vicinity of criticality,
indicate rough idea about the mass and decay constant of TD as follows.
Substituting our numerical result, Eq.~(\ref{vac-num-2loop}),  
into  Eq.~(\ref{md}), we find 
\begin{equation}
  M_{\rm TD}^2 \simeq 3.02 \, \eta \frac{m^4}{F_{\rm TD}^2} \,.
  \label{massestimate}
\end{equation}

Furthermore, by using $F_\pi/m \simeq 0.41 \sqrt{N_D N_{\rm TC}/3}$ 
in Eq.~(\ref{PS}), we obtain
\begin{equation}
 \frac{M_{\rm TD}}{F_\pi} \simeq 
 3.5 \frac{F_\pi}{F_{\rm TD}} \cdot
 \left(\sqrt{\frac{N_f}{2N_D}} \sqrt{\frac{8}{2N_D}}
 \sqrt{\frac{2}{N_{\rm TC}}} \right)  \, .
\end{equation}
The TD with mass, say $M_{\rm TD} \sim $ 500~GeV, 
would require TD coupling smaller than that of the  Standard Model Higgs 
by $F_\pi/F_{\rm TD} \simeq  3/5$ up to model-dependent factors of 
$N_f, N_D$ and $N_{\rm TC}$ besides other dynamical details. 
If the TD mass is much smaller, $F_{\rm TD} \ll F_\pi$, 
on the other hand, it could lead to a decoupled TD, 
which might be a candidate  of dark matter.
Detailed studies are required in order to confirm whether or not
such a decoupled TD satisfies conditions for dark matter.

\section{Summary and discussions}

We have studied analytically the improved ladder SD equation
with the parabolic approximation for the beta function
and also analyzed numerically the solution of the ladder
SD equation with the two-loop exact running gauge coupling.

We explicitly calculated the techni-gluon condensate near the conformal edge
and found that the behavior is
$\VEV{G_{\mu\nu}^2}/m^4 \sim (\alpha/\alpha_{\rm cr}-1)^{-3/2} \to \infty$
($\alpha \to \alpha_{\rm cr}$) with $m\ll \Lambda_{\rm TC}$, 
in accord with Ref.~\cite{Haba:2010iv} 
but in disagreement with the assumption of 
Refs.~\cite{Bando:1986bg, Appelquist:2010gy}. 
The numerical calculation is consistent with this analytic result.
This situation is different from QCD having 
no approximate conformal symmetry, where $\Lambda_{\rm QCD} \sim m$ and 
$\VEV{G_{\mu\nu}^2}_{\rm full}  \sim \VEV{G_{\mu\nu}^2}_{\rm perturbative} 
 \sim \Lambda_{\rm QCD}^4 $.

On the other hand, the vacuum energy (divided by $m^4$) is finite, 
$V/m^4 \rightarrow {\rm const}$,
even in the critical limit, as in the case of the fixed gauge coupling.
Our result for the vacuum energy only yields a combination of 
the mass  $M_{\rm TD}$ and the decay constant $F_{\rm TD}$ 
through PCDC but not each of them separately, 
as was the case for most of the discussions in the literature.
Combining the PCDC relation, Eq. (\ref{md}), 
$F_{\rm TD}^2 M_{\rm TD}^2 = - 4 \VEV{\theta_\mu^\mu}$,
with the numerical result for the vacuum energy, Eq. (\ref{vac-num-2loop}), 
we found $M_{\rm TD}^2 = 3.02 \, \eta \, m^4/F_{\rm TD}^2$
with $\eta \equiv N_{\rm TC} N_f/(2\pi^2)$, Eq.(\ref{massestimate}).
This relation implies $M_{\rm TD}/m \sim m/F_{\rm TD}$ near 
the conformal edge and hence naturally $M_{\rm TD} ={\cal O} (m)$ 
in contrast to Refs.~\cite{Appelquist:2010gy,Dietrich:2005jn}. 
(A similar conclusion was made in a different context~\cite{Vecchi:2010aj}.)
As an extreme possibility  we could have 
$M_{\rm TD}/m \rightarrow 0$ only when   $m/F_{\rm TD}\rightarrow 0$ 
and the TD gets decoupled. 
If such an idealized  decoupled massless TD is realized 
at the conformal edge, the light decoupled TD as a pseudo NG boson 
slightly off the conformal edge may be a candidate for the dark matter.

The scale anomaly formally yields the relation
$\beta/(4\alpha) = \VEV{\theta_\mu^\mu}/\VEV{G_{\mu\nu}^2}$,
so that the above results imply the nonperturbative behavior
of the beta function reflecting the nonperturbative effects
of the dynamical mass generation,
$\beta(\alpha) \sim - (\alpha/\alpha_{\rm cr}-1)^{3/2}$.
Numerically, we obtained 
$\beta/(4\alpha) = \xi \, (\alpha/\alpha_{\rm cr}-1)^{3/2}/\lambda_*$
with $\xi = -0.026$ for the two-loop running gauge coupling.
The absolute value of the coefficient is smaller than that 
for the fixed gauge coupling, $\xi = -0.04$.
However, the exponent $\kappa$ of the nonperturbative beta function
at the conformal edge seems universal, $\kappa = 3$,
where $\beta/(4\alpha) \propto (\alpha/\alpha_{\rm cr}-1)^{\kappa/2}$.
This nature of the nonperturbative beta function having
the multiple zero is crucial to reproduce the essential singularity scaling
Eq.~(\ref{ess-sing}).
 
We have settled some of the controversy related with the TD mass 
raised within the improved-ladder SD equation. 
However, several issues remain to be explored:

In particular, a central problem is how large 
the TD mass $M_{\rm TD}$ is.
In order to discuss collider phenomenology of the TD and 
also check whether or not a decoupled TD is in fact realized 
near the conformal edge,
we should obtain mass $M_{\rm TD}$ and the decay constant $F_{\rm TD}$ separately.

Thus we would need more information other than the vacuum energy. 
As was mentioned in the Introduction, such a calculation was in fact 
done in a most straightforward way~\cite{Harada:2003dc}, based on 
the SD and BS equations in the improved ladder approximation with 
the two-loop running coupling constant having the CBZ IR fixed point, 
which suggests $M_{\rm TD} \sim \sqrt{2}m$, Eq.(\ref{SDBS}), 
without evidence of the  decoupled light TD much smaller than $m$. 
Note however that this calculation was actually done only numerically 
and at slightly off the conformal edge, and hence the result is not 
conclusive about the very close point to the conformal edge. 

On the other hand,  in the holographic framework~\cite{Haba:2010iv} 
which has a wider parameter space than that of the (improved) 
ladder approximation so as to adjust the $S$ parameter arbitrarily small,
it was shown that at the limit of conformal edge 
$m/\Lambda_{\rm TC} \rightarrow 0$ the techni-gluon condensate 
vanishes $\Gamma \rightarrow \infty$, 
with $\Gamma$  parameterized as in Eq.(\ref{Gamma}), 
which in turn implies $M_{\rm TD}/m \rightarrow 0$ at the sacrifice of
decoupling $m/F_{\rm TD} \rightarrow 0$, although such an extreme case 
is unlikely for the realistic setting of the typical TC model building
(slightly away from the conformal edge 
$m/\Lambda_{\rm TC} = 10^{-3}-10^{-4}$) where the holography suggests 
$M_{\rm TD}/m ={\cal O} (1)$, or 
$M_{\rm TD} \sim 600 \, {\rm  GeV}$ (Eq.(\ref{TD2})).
So although the theoretical possibility for the decoupled TD at 
the conformal edge is not completely excluded, 
there is no signature of such a possibility at least 
in near conformal edge region relevant to  the realistic TC model building.

We have not included  interactions like ETC (Extended Technicolor) 
between the techni-fermions and the quarks/leptons which should be 
included to give mass to the quarks and leptons in the realistic TC models.  
Including these interactions also induce additional interactions 
among the techni-fermions themselves, which may be described by 
the effective four-fermion interactions in addition to 
the TC gauge interactions  we have discussed (``gauged NJL model''). 
Such ETC-like effects on the TD mass were already studied intensively 
in the ladder SD equation with nonrunning gauge 
coupling~\cite{Bardeen:1985sm,Kondo:1988qd,Nonoyama:1989dq}, 
with the results $M_{\rm TD} \sim {\cal O} (m)$, i.e., 
against very light TD mass and decoupled TD, as was mentioned 
in the Introduction.
As is clear from our arguments in this paper, 
the situation with the additional four-fermion interaction 
in the ladder approximation with nonrunning gauge coupling 
will remain  essentially the same in the improved ladder SD equation
with the two-loop running gauge coupling. 
Moreover, more elaborate calculations in the gauged NJL 
model~\cite{Shuto:1989te} showed that 
$M_{\rm TD} \rightarrow \sqrt{2} m$ (Eq.(\ref{gaugedNJLTD})) 
in such a limit  along the whole critical line 
($0< \alpha\leq \alpha_{\rm cr}$). 
Note also that the result Eq.(\ref{gaugedNJLTD})~\cite{Shuto:1989te} 
is consistent with Eq.(\ref{SDBS})~\cite{Harada:2003dc} 
which is the straightforward computation of the spectra within 
the ladder SD and BS equations in the improved ladder approximation with
the two-loop gauge coupling near the conformal edge 
(without four-fermion coupling).  
Note however that these calculations did only for
the inverse propagator at  zero-momentum but not the pole mass 
(on-shell mass), and hence are still not conclusive.
We definitely need more reliable calculations such as 
the lattice simulations about the TD spectrum.  

We have considered TD as a bound state of techni-fermion and 
anti-techni-fermion both of which acquired mass $m$. 
The mass $m$ spontaneously breaks chiral symmetry and 
at the same time breaks spontaneously and explicitly the scale symmetry,
the scale anomaly due to this mass generation being of order 
${\cal O} (m^4)$ as we computed explicitly in this paper.
Hence such a bound state should  have mass of order 
${\cal O} (m) (\ll \Lambda_{\rm TC})$.
On the contrary, it was argued~\cite{Haba:2010iv} 
that the techni-glueball mass should be of order 
${\cal O} (\Lambda_{\rm TC})$, since the scale-symmetry breaking 
free from  the techni-fermion mass generation is due to 
the scale anomaly of order ${\cal O} (\Lambda_{\rm TC}^4)$ 
corresponding to the usual perturbative running of the coupling 
for $\mu > \Lambda_{\rm TC}$ ($\langle \theta_\mu^\mu\rangle_{\rm perturbative}$
in Eq.(\ref{NPdef}) ).
Then we expect little mixing between our TD and the techni-glueball,
in sharp contrast to QCD where the flavor-single scalar meson 
(analogue of TD) and the scalar glueball may mix strongly. 
More reliable calculations  are of course highly desired. 

In this paper, we assumed that the fermion loop is dominant
in the techni-gluon condensate and the vacuum energy.
In principle, there might exist nonperturbative techni-gluonic effects.
It is difficult to estimate it in our approach, however.
A lattice simulation may also resolve this issue.
We shed a light on the problem which has made confusion 
in the improved ladder approximation with the two-loop running gauge coupling.
We clarified it in the analytical and numerical ways 
within the same framework of the improved ladder SD approximation.
There should exist nonperturbative effects beyond the improved
ladder approximation,
although it is very unclear whether or not they are relevant.
Toward a conclusive answer, one might challenge a more rigorous approach
such as a lattice gauge theory.
We hope that our results will be reconfirmed more rigorously in future.

Needless to say, our analysis in this paper is also applicable to 
dynamical symmetry breaking scenarios with 
large anomalous dimension/conformality other than TC,
such as the top-mode standard model with extra dimensions 
which has a UV fixed point~\cite{Hashimoto:2000uk},  
higher representation quark condensate model~\cite{Marciano:1980zf},  
or even the QCD with finite temperature where the running of 
the coupling will be frozen (mocking conformal) in 
the IR region below the temperature scale, etc.. 

\acknowledgments

The authors thank T.~Appelquist and T.~Kugo for fruitful discussions.
M.H. was supported by Maskawa Institute for Science and Culture, 
Kyoto Sangyo University.
This work was supported in part by the JSPS Grant-in-Aid for 
Scientific Research(S) \# 22224003.

\end{document}